\documentclass{aa}
\newcommand\gsim{~\lower.5ex\hbox{$\buildrel > \over \sim$}~}

\defcitealias{Dijkstra2010}{DW10}

\usepackage[varg]{txfonts}
\usepackage[colorlinks=true,linkcolor=blue,allcolors=blue]{hyperref}
\begin{document} 

\title{Predicting Ly$\alpha$ escape fractions with a simple observable\thanks{Based on observations obtained with the Very Large Telescope, programs: 098.A-0819 \& 099.A-0254.}}
\subtitle{Ly$\alpha$ in emission as an empirically calibrated star formation rate indicator}
\author{David Sobral\inst{1,2}\thanks{\email{d.sobral@lancaster.ac.uk}}  \and Jorryt Matthee\inst{2,3}}

\institute{Department of Physics, Lancaster University, Lancaster, LA1 4YB, UK
\and  Leiden Observatory, Leiden University, P.O.\ Box 9513, NL-2300 RA Leiden, The Netherlands
\and Department of Physics, ETH Z\"urich, Wolfgang-Pauli-Strasse 27, 8093 Z\"urich, Switzerland  }

%\date{Received ...; accepted ...}
\authorrunning{Sobral \& Matthee}

\abstract{Lyman-$\alpha$ (Ly$\alpha$) is intrinsically the brightest line emitted from active galaxies. While it originates from many physical processes, for star-forming galaxies the intrinsic Ly$\alpha$ luminosity is a direct tracer of the Lyman-continuum (LyC) radiation produced by the most massive O- and early-type B-stars ($M_{\star}\gsim10$\,M$_{\odot}$) with lifetimes of a few Myrs. As such, Ly$\alpha$ luminosity should be an excellent instantaneous star formation rate (SFR) indicator. However, its resonant nature and susceptibility to dust as a rest-frame UV photon makes Ly$\alpha$ very hard to interpret due to the uncertain Ly$\alpha$ escape fraction, $\rm f_{esc,Ly\alpha}$. Here we explore results from the CAlibrating LYMan-$\alpha$ with H$\alpha$ (CALYMHA) survey at $z=2.2$, follow-up of Ly$\alpha$ emitters (LAEs) at $z=2.2-2.6$ and a $z\sim0-0.3$ compilation of LAEs to directly measure $\rm f_{esc,Ly\alpha}$ with H$\alpha$. We derive a simple empirical relation that robustly retrieves $\rm f_{esc,Ly\alpha}$ as a function of Ly$\alpha$ rest-frame EW (EW$_0$): $\rm f_{esc,Ly\alpha}=0.0048\,EW_0[{\AA}]\pm0.05$ and we show that it constrains a well-defined anti-correlation between ionisation efficiency ($\rm \xi_{\rm ion}$) and dust extinction in LAEs. Observed Ly$\alpha$ luminosities and EW$_0$ are easy measurable quantities at high redshift, thus making our relation a practical tool to estimate intrinsic Ly$\alpha$ and LyC luminosities under well controlled and simple assumptions. Our results allow observed Ly$\alpha$ luminosities to be used to compute SFRs for LAEs at $z\sim0-2.6$ within $\pm0.2$\,dex of the H$\alpha$ dust corrected SFRs. We apply our empirical SFR(Ly$\alpha$,EW$_0$) calibration to several sources at $z\geq2.6$ to find that star-forming LAEs have SFRs typically ranging from 0.1 to 20\,M$_{\odot}$\,yr$^{-1}$ and that our calibration might be even applicable for the most luminous LAEs within the epoch of re-ionisation. Our results imply high ionisation efficiencies ($\rm \log_{10}[\xi_{ion}/Hz\,erg^{-1}]=25.4-25.6$) and low dust content in LAEs across cosmic time, and will be easily tested with future observations with {\it JWST} which can obtain H$\alpha$ and H$\beta$ measurements for high-redshift LAEs.}


\maketitle

%%%%%%%%%%%%%%%%%%%    Introduction     %%%%%%%%%%%%%%%%%
\section{Introduction}

With a vacuum rest-frame wavelength of 1215.67\,{\AA}, the Lyman-$\alpha$ (Ly$\alpha$) recombination line ($n=2 \rightarrow n=1$) plays a key role in the energy release from ionised hydrogen gas, being intrinsically the strongest emission line in the rest-frame UV and optical \citep[e.g.][]{Partridge1967,Pritchet1994}. Ly$\alpha$ is emitted from ionised gas around star-forming regions \citep[e.g.][]{Charlot1993,Pritchet1994} and AGN \citep[e.g.][]{Miley2008} and it is routinely used as a way to find high redshift sources ($z\sim2-7$; see e.g. \citealt{Malhotra2004}). 

Several searches for Ly$\alpha$-emitting sources (Ly$\alpha$ emitters; LAEs) have led to samples of thousands of star-forming galaxies (SFGs) and AGN \citep[e.g.][and references therein]{Sobral2017_SC4K}. LAEs are typically faint in the rest-frame UV, including many that are too faint to be detected by continuum based searches even with the {\it Hubble Space Telescope} \citep[e.g.][]{Bacon2015}. The techniques used to detect LAEs include narrow-band surveys \citep[e.g.][]{Rhoads2000,Ouchi2008,Hu2010, Matthee2015}, Integral Field Unit (IFU) surveys \citep[e.g.][]{vanBreukelen2005,Drake2017a} and blind slit spectroscopy \citep[e.g.][]{Martin2004,Rauch2008,Cassata2011}. Galaxies selected through their Ly$\alpha$ emission allow for easy spectroscopic follow-up due to their high EWs \citep[e.g.][]{Hashimoto2017} and typically probe low stellar masses \citep[see e.g.][]{Gawiser2007,Hagen2016}.

The intrinsic Ly$\alpha$ luminosity is a direct tracer of the ionising Lyman-continuum (LyC) luminosity and thus a tracer of instantaneous star formation rate (SFR), in the same way as H$\alpha$ is \citep[e.g.][]{Kennicutt1998}. Unfortunately, inferring intrinsic properties of galaxies from Ly$\alpha$ observations is extremely challenging. This is due to the complex resonant nature and sensitivity to dust of Ly$\alpha$ \citep[see e.g.][for a detailed review on Ly$\alpha$]{Dijkstra2017}, which contrasts with H$\alpha$. For example, a significant fraction of Ly$\alpha$ photons is scattered in the Inter-Stellar Medium (ISM) and in the Circum-Galactic Medium (CGM) as evidenced by the presence of extended Ly$\alpha$ halos in LAEs \citep[e.g.][]{Momose2014,Wisotzki2016}, but also in the more general population of $z\sim2$ SFGs sampled by H$\alpha$ emitters \citep[][]{Matthee2016}, and the bluer component of such population traced by UV-continuum selected galaxies \citep[e.g.][]{Steidel2011}. Such scattering leads to kpc-long random-walks which take millions of years and that significantly increase the probability of Ly$\alpha$ photons being absorbed by dust particles. The complex scattering and consequent higher susceptibility to dust absorption typically leads to low and uncertain Ly$\alpha$ escape fractions \citep[$\rm f_{esc,Ly\alpha}$; the ratio between observed and intrinsic Ly$\alpha$ luminosity; see e.g.][]{Atek2008}.

``Typical" star-forming galaxies at $z\sim2$ have low $\rm f_{esc,Ly\alpha}$ ($\sim1-5$\%; e.g. \citealt{Oteo2015,Cassata2015}), likely because significant amounts of dust present in their ISM easily absorb Ly$\alpha$ photons \citep[e.g.][]{Ciardullo2014,Oteo2015,Oyarzun2017}. However, sources selected through their Ly$\alpha$ emission typically have $\sim10$ times higher $\rm f_{esc,Ly\alpha}$ \citep[e.g.][]{Song2014,Sobral2017}, with Ly$\alpha$ escaping over $\approx2\times$ larger radii than H$\alpha$ \citep[][]{Sobral2017}.

Furthermore, one expects $\rm f_{esc,Ly\alpha}$ to depend on several physical properties which could be used as predictors of $\rm f_{esc,Ly\alpha}$. For example, $\rm f_{esc,Ly\alpha}$ anti-correlates with stellar mass \citep[e.g.][]{Oyarzun2017}, dust attenuation \citep[e.g.][]{Verhamme2008,Hayes2011,Matthee2016, An2017} and SFR \citep[e.g.][]{Matthee2016}. However, most of these relations require derived properties \citep[e.g.][]{Yang2017}, show a large scatter, may evolve with redshift and sometimes reveal complicated trends \citep[e.g. dust dependence; see][]{Matthee2016}.

Interestingly, the Ly$\alpha$ rest-frame equivalent width (EW$_0$), a simple observable, seems to be the simplest direct predictor of $\rm f_{esc,Ly\alpha}$ in LAEs \citep[][]{Sobral2017,Verhamme2017} with a relation that shows no strong evolution from $z\sim0$ to $z\sim2$ \citep[][]{Sobral2017} and that might be applicable at least up to $z\sim5$ \citep[][]{Harikane2017}. Such empirical relation may hold the key for a simple but useful calibration of Ly$\alpha$ as a direct tracer of the intrinsic LyC luminosity \citep[see][and references therein]{Reddy2016,Steidel2018,Fletcher2018} by providing a way to estimate $\rm f_{esc,Ly\alpha}$, and thus as a good SFR indicator for LAEs \citep[see also][hereafter \citetalias{Dijkstra2010}]{Dijkstra2010}. We fully explore such possibility and its implications in this work. Note that this paper makes no attempt to simplify the complex radiative transfer by which Ly$\alpha$ photons escape from galaxies. Instead, this work focuses on an empirical approach to predict Ly$\alpha$ escape fractions with a simple observable based on direct observations. In \S\ref{Sample_methods} we present the samples at different redshifts and methods used to compute $\rm f_{esc,Ly\alpha}$. In \S\ref{results_discc} we present and discuss the results, their physical interpretation and our proposed empirical calibration of Ly$\alpha$ as an SFR indicator. Finally, we present the conclusions in \S\ref{conclusions}. We use AB magnitudes \citep[][]{Oke1983}, a \cite{Salpeter1955} initial mass function (IMF; with mass limits 0.1 and 100\,M$_{\odot}$) and adopt a flat cosmology with $\Omega_\mathrm{m}=0.3$, $\Omega_\Lambda=0.7$, and $H_0=70$\,km\,s$^{-1}$\,Mpc$^{-1}$. 

\section{Sample and Methods}\label{Sample_methods}

In this study we use a large compilation of LAEs which have been widely studied in the literature \citep[e.g.][]{Cardamone2009,Henry2015,Trainor2015,Verhamme2017,Sobral2017} at $z\leq0.3$ and $z\sim2.2-2.6$ with measured or inferred dust-extinction corrected H$\alpha$ luminosities and thus $\rm f_{esc,Ly\alpha}$ available. We note that these cover sources from low ($\approx5$\,{\AA}) to high ($\approx160$\,{\AA}) EW$_0$ across a range of redshifts, with SFRs typically around $\sim5-50$\,M$_{\odot}$\,yr$^{-1}$ (typical of LAEs) at $z\sim0.3-2.6$. The sample combines sources obtained with somewhat heterogeneous selections which allow us to obtain a more conservative scatter in the trends we investigate. Our approach also allows us to obtain relations that are more widely applicable for LAEs with measured Ly$\alpha$ luminosities and EW$_0$. We note nonetheless that our results are only valid for LAEs and are empirically based on observables. Note that in this study we explore luminosities within $\approx2-3$\,arcsec (typically $\approx15-20$\,kpc) diameters. These do not explicitly include the even more extended Ly$\alpha$ halo luminosity beyond $\sim20$\,kpc, but we refer interested readers to studies that have investigated the spatial dependence of the Ly$\alpha$ escape fraction for different sources \citep[e.g.][]{Matthee2016,Sobral2017}.

\subsection{LAEs at low redshift ($z\leq0.3$)}

For our lower redshift sample, we explore a compilation of 30 sources presented in \cite{Verhamme2017} which have accurate (H$\alpha$ derived) $\rm f_{esc,Ly\alpha}$ measurements and sample a range of galaxy properties. The sample includes high EW H$\alpha$ emitters (HAEs) from the Lyman Alpha Reference Sample at $z=0.02-0.2$ \citep[LARS, e.g.][]{Hayes2013,Hayes2014}, a sample of LyC leakers (LyCLs) investigated in \cite{Verhamme2017} at $z\sim0.3$ \citep[][]{Izotov2016a,Izotov2016b} and a more general `green pea' (GPs) sample \citep[e.g.][]{Cardamone2009,Henry2015,Yang2016,Yang2017}. These are all LAEs at low redshift with available Ly$\alpha$, H$\alpha$ and dust extinction information required to estimate $\rm f_{esc,Ly\alpha}$ (see \S \ref{Measuring_fesc}) and for which Ly$\alpha$ EW$_0$s are available. For more details on the sample, see \cite{Verhamme2017} and references therein.

\subsection{LAEs at cosmic noon ($z=2.2-2.6$)}

For our sample at the peak of star formation history we use 188 narrow-band selected LAEs with H$\alpha$ measurements from the CALYMHA survey at $z=2.2$ \citep[][]{Matthee2016,Sobral2017} presented in \cite{Sobral2017}, for which $\rm f_{esc,Ly\alpha}$ measurements are provided as a function of EW$_0$. In addition, we explore spectroscopic follow-up of CALYMHA sources with X-SHOOTER on the VLT \citep[][]{Sobral2018b} and individual measurements for four sources (CALYMHA-67, -93, -147 and -373; see \citealt{Sobral2018b}). For those sources we measure Ly$\alpha$, H$\alpha$ and H$\beta$ and correct for dust extinction as in \S\ref{Measuring_fesc}.

Furthermore, we also use a sample of 29 narrow-band selected LAEs at $z\sim2.6$ presented by \cite{Trainor2015}, for which Ly$\alpha$ and H$\alpha$ measurements are available. We use results from \cite{Trainor2016} that show that for the full sample the Balmer decrement is consistent with $\approx0$\,mag of extinction. This is dominated by the more numerous sources with higher EWs, and thus we assume $\approx0$\,mag ($\rm A_{\rm H\alpha}$) of extinction for the highest EW bin. For the sources with the lowest EWs, we correct for $\rm A_{\rm H\alpha}=0.1$\,mag of extinction, as these are the most massive sources and thus expected to be slightly more dusty \citep[see][]{GarnBest2010}. We note that our obscuration correction may be a slight underestimation (resulting in over-estimating the escape fraction at the lowest EWs) for the \cite{Trainor2015} sample.

\subsection{Higher redshift LAEs ($2.6\leq z\leq6$)}

As an application of our results, we explore the publicly available sample of 3,908 LAEs in the COSMOS field \citep[SC4K survey;][]{Sobral2017_SC4K} which provides Ly$\alpha$ luminosities and rest-frame EWs for all LAEs. We also explore published median or average values for the latest MUSE samples, containing 417 LAEs \citep[e.g.][]{Hashimoto2017}. Note that for all these higher redshift samples, H$\alpha$ is not directly available, thus $\rm f_{esc,Ly\alpha}$ cannot be directly measured \citep[but see][]{Harikane2017}.

%%%%%%%%%%%%%%%%%%%%%%%%%%%%%%%%%%%%%%%%%%%%
%
% FIGURE 1 - Relation between Lya fesc and Lya EW_0, fits and comparisons
%
%%%%%%%%%%%%%%%%%%%%%%%%%%%%%%%%%%%%%%%%%%%%
\begin{figure*}
\centering
\includegraphics[width=14.0cm]{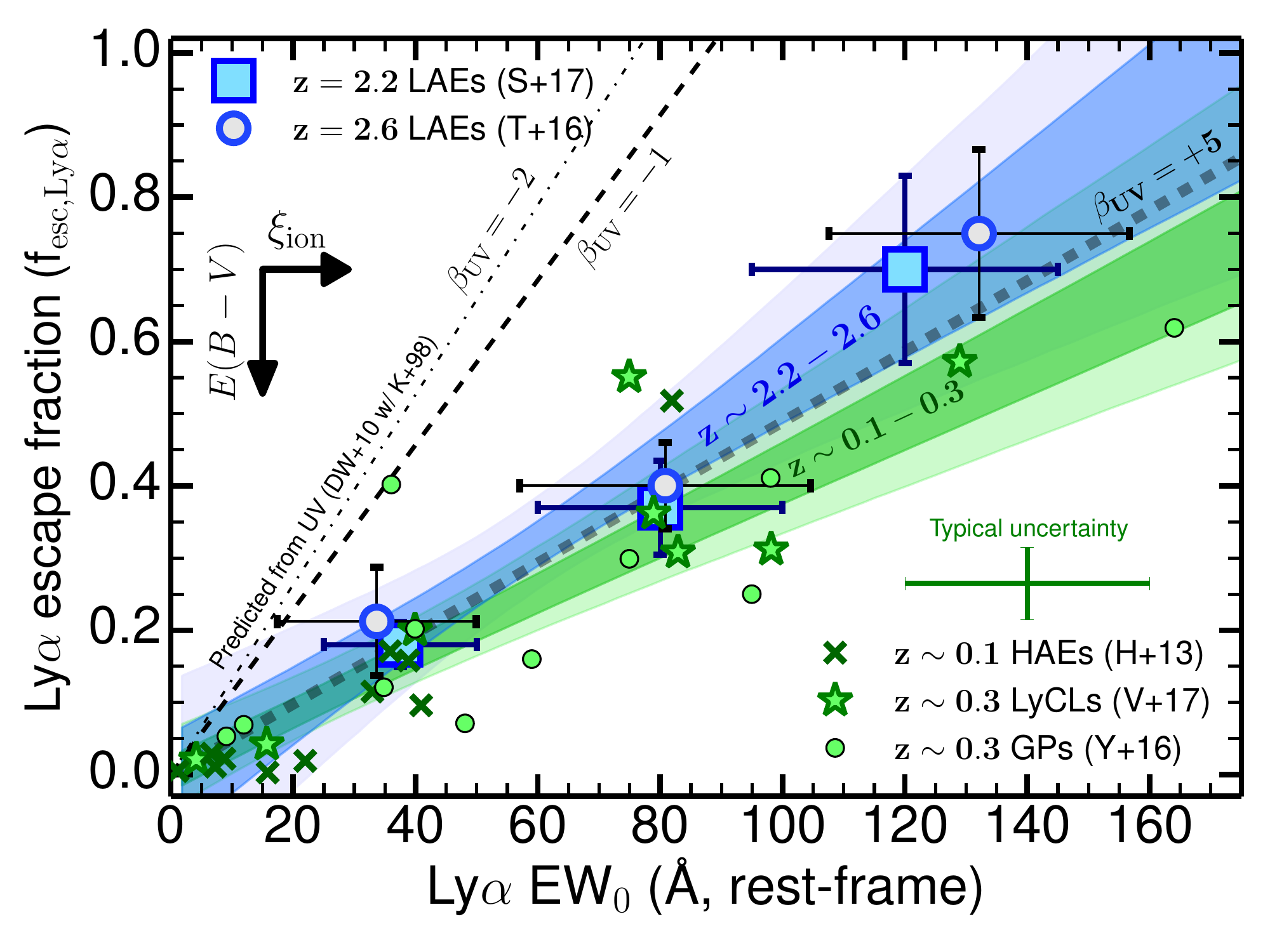}
\caption{The relation between $\rm f_{esc,Ly\alpha}$ and Ly$\alpha$ EW$_0$ for $z\sim2.2$ \citep[stacks; see][]{Sobral2017}, $z\sim2.6$ \citep[binning;][]{Trainor2015} and comparison with $z\sim0-0.3$ samples \citep[e.g.][]{Cardamone2009,Hayes2013,Henry2015,Yang2016,Yang2017,Verhamme2017}, estimated from dust-corrected H$\alpha$ luminosities (Equation \ref{eq1}). We show the 1\,$\sigma$ and 2\,$\sigma$ range for the fits at $z\sim2.2-2.6$ and $z\sim0-0.3$  separately, and find them to be consistent within those uncertainties, albeit with a potential steeper relation at higher redshift. We find a combined best fitting relation given by $\rm f_{esc,Ly\alpha}=0.0048\,EW_{0}\pm0.05$. The observed relation is significantly away from what would be predicted based on observed UV slopes between $\beta\approx-2$ and $\beta\approx-1$ for LAEs \citepalias[see][]{Dijkstra2010} and would require $\beta\approx+5$ for a good fit using Equation \ref{final_DW10_s}. Such red $\beta$ slopes are not observed for LAEs. Modifying Equation \ref{final_DW10_s} to include the effect of $\rm \xi_{\rm ion}$ and dust reveals that those physical parameters likely play an important role; see Sections \ref{predict_from_UV} and \ref{phy_interp}.}
\label{EW_vs_f_esc}
\end{figure*}  
%%%%%%%%%%%%%%%%%%%%%%%%%%%%

\subsection{Measuring the Ly$\alpha$ escape fraction ($f_{esc,Ly\alpha}$) with H$\alpha$} \label{Measuring_fesc}

We use dust corrected H$\alpha$ luminosity to predict the intrinsic Ly$\alpha$ luminosity. We then compare the latter to the observed Ly$\alpha$ luminosity to obtain the Ly$\alpha$ escape fraction ($\rm f_{esc,Ly\alpha}$). Assuming case B recombination\footnote{We use Ly$\alpha$/H$\alpha=8.7$, but vary the Ly$\alpha$/H$\alpha$ case B ratio between 8.0 and 9.0 to test for its effect; see \S\ref{emp_SFR_err} and also discussions in \cite{Henry2015}.}, a temperature of 10$^4$\,K and an electron density of 350\,cm$^{-3}$, we can use the observed Ly$\alpha$ luminosity ($\rm L_{Ly\alpha}$), the observed H$\alpha$ luminosity ($\rm L_{H\alpha}$) and the dust extinction affecting $\rm L_{H\alpha}$ ($\rm A_{\rm H\alpha}$\footnote{With our case B assumptions the intrinsic Balmer decrement is: $\rm H\alpha/H\beta=2.86$. Using a \citet{Calzetti00} dust attenuation law we use $\rm A_{\rm H\alpha}=\rm 6.531\log_{10}(H\alpha/H\beta)-2.981$ \citep[see details in e.g.][]{Sobral2012}.}, in mag) to compute $\rm f_{esc,Ly\alpha}$ as:
\begin{equation}\label{eq1}
f_{\rm esc,Ly\alpha} =  \frac{\rm L_{\rm Ly\alpha}}{\rm 8.7\,L_{\rm H\alpha} \times 10^{0.4\times A_{\rm H\alpha}}}.
\end{equation}
This means that with our assumptions so far, and provided that we know $\rm f_{esc,Ly\alpha}$, we can use the observed $\rm L_{Ly\alpha}$ to obtain the intrinsic H$\alpha$ luminosity. All sources or samples in this study have been corrected for dust extinction using Balmer decrements, either measured directly for individual sources, or by applying the median extinction for stacks or bins of sources. Therefore, one can use Ly$\alpha$ as a star formation rate (SFR) indicator\footnote{For continuous star formation over 10\,Myr timescales and calibrated for solar metallicity; see \citet{Kennicutt1998}.} following \citet{Kennicutt1998} for a Salpeter (Chabrier) IMF $(0.1-100 \,{\rm M}_{\sun})$:
\begin{equation} \label{eq:SFR}
\rm {\rm SFR_{Ly\alpha}\,[M_{\odot}\,yr^{-1}]}=\frac{7.9 (4.4)\times10^{-42}}{(1-f_{esc,LyC})} \, \, \frac{L_{\rm Ly\alpha}}{8.7\,f_{\rm esc,Ly\alpha}}
\end{equation}
where $\rm f_{esc,LyC}$ is the escape fraction of ionising LyC photons \citep[see e.g.][]{Sobral2017_SC4K}. In practice, $\rm f_{esc,LyC}$ is typically assumed to be $\approx0$, but it may be $\approx0.1-0.15$ for LAEs \citep[see discussions in e.g.][]{MattheeGALEX2017,Verhamme2017}.

\subsection{Statistical fits and errors}\label{stat_fits}

For all fits and relations in this work (e.g. $\rm f_{esc,Ly\alpha}$ vs. EW$_0$), we vary each data-point or binned data-point within its full Gaussian probability distribution function independently (both in EW$_0$ and $\rm f_{esc,Ly\alpha}$), and re-fit 10,000 times. We present the best-fit relation as the median of all fits, and the uncertainties (lower and upper) are the 16 and 84 percentiles. For bootstrapped quantities (e.g. for fitting the low redshift sample) we obtain 10,000 samples randomly picking half of the total number of sources and computing that specific quantity. We fit relations in the form $y=Ax+B$.

\section{Results and Discussion}\label{results_discc}

\subsection{The observed $\rm f_{esc,Ly\alpha}$-EW$_0$ relation at $z\sim0.1-2.6$}\label{fits_empiri}

Figure \ref{EW_vs_f_esc} shows that $\rm f_{esc,Ly\alpha}$ correlates with Ly$\alpha$ EW$_{0}$ with apparently no redshift evolution between $z=0-2.6$ \citep[see also][]{Verhamme2017,Sobral2017}. We find that $\rm f_{esc,Ly\alpha}$ varies continuously from $\approx0.2$ to $\approx0.7$ for LAEs from the lowest ($\approx30$\,\AA) to the highest ($\approx120-160$\,\AA) Ly$\alpha$ rest-frame EWs. We use our samples at $z\sim0-0.3$ and $z\sim2.2-2.6$, separately and together, to obtain linear fits to the relation between $\rm f_{esc,Ly\alpha}$ and Ly$\alpha$ EW$_{0}$ (see \S\ref{stat_fits}). These fits allow us to provide a more quantitative view on the empirical relation and evaluate any subtle redshift evolution; see Table \ref{Fits_fesc_EW}.

The relation between $\rm f_{esc,Ly\alpha}$ and Ly$\alpha$ EW$_{0}$ is statistically significant at 5 to 10\,$\sigma$ for all redshifts. We note that all linear fits are consistent with a zero escape fraction for a null EW$_0$ (Table \ref{Fits_fesc_EW}), suggesting that the trend is well extrapolated for weak LAEs with EW$_0\approx0-20$\,{\AA}. Furthermore, as Table \ref{Fits_fesc_EW} shows, the fits to the individual (perturbed) samples at different redshifts result in relatively similar slopes and normalisations within the uncertainties, and thus are consistent with the same relation from $z\sim0$ to $z\sim2.6$. Nevertheless, we note that there is minor evidence for a shallower relation at lower redshift for the highest EW$_0$ (Figure \ref{EW_vs_f_esc}), but this could be driven by current samples selecting sources with more extreme properties (including LyC leakers). Given our findings, we decide to combine the samples and obtain joint fits, with the results shown in Table \ref{Fits_fesc_EW}. The slope of the relation is consistent with being $\approx0.005$ with a null $\rm f_{esc,Ly\alpha}$ for EW$_{0}=0$\,{\AA}.

%%%%%%%%%%%%%%%%%%%%%%%%%%%%%%%%%%%%%%%%%%%
%
% Table 1 - Fits to the fesc,Lya vs EW_0 relation for the various sub-samples
%
%%%%%%%%%%%%%%%%%%%%%%%%%%%%%%%%%%%%%%%%%%%
\begin{table}
 \centering
\caption{The results from fitting the relation between $\rm f_{esc,Ly\alpha}$ and Ly$\alpha$ EW$_0$ as $\rm f_{esc,Ly\alpha}=A\times {\rm EW_0}+B$, with EW$_0$ in {\AA} (see \S\ref{stat_fits}). [i: individual sources used for fitting; b: binned/averaged quantity used for fitting; B: bootstrap analysis when fitting each of the 10,000 times; G: each data bin is perturbed along its Gaussian probability distribution.]}
 \label{Fits_fesc_EW}
\begin{tabular}{ccccc}
\hline
Sample   & A (\AA$^{-1}$)  & B & [notes] & \\ 
\hline
$z\sim0-0.3$  &  $0.0041^{+0.0006}_{-0.0004}$ & $0.00^{+0.03}_{-0.02}$  & [i,B] \\ 
$z\sim2.2$  &  $0.0056^{+0.0012}_{-0.0011}$ & $0.00^{+0.05}_{-0.05}$  & [b,G] \\ 
$z\sim2.6$  &  $0.0054^{+0.0016}_{-0.0015}$ & $0.01^{+0.11}_{-0.11}$  & [b,G] \\ 
\hline
$z\sim0-2.2$  &  $0.0045^{+0.0008}_{-0.0007}$ & $0.00^{+0.06}_{-0.06}$  & [b,G] \\ 
$z\sim2.2-2.6$  &  $0.0056^{+0.0012}_{-0.0012}$ & $0.00^{+0.07}_{-0.08}$  & [b,G] \\ 
\hline
$z\sim0-2.6$  &  $0.0048^{+0.0007}_{-0.0007}$ & $0.00^{+0.05}_{-0.05}$  & [b,G] \\ 
\hline
\end{tabular}
\end{table}

\subsection{The $\rm f_{esc,Ly\alpha}$-EW$_0$ relation: expectation vs. reality}\label{predict_from_UV}

The existence of a relation between $\rm f_{esc,Ly\alpha}$ and EW$_0$ (Figure \ref{EW_vs_f_esc}) is not surprising. This is because Ly$\alpha$ EW$_{0}$ is sensitive to the ratio between Ly$\alpha$ and the UV luminosities, which can be used as a proxy of $\rm f_{esc,Ly\alpha}$ \citep[see e.g.][]{Dijkstra2010,Sobral2017_SC4K}. However, the slope, normalisation and scatter of such relation depend on complex physical conditions such as dust obscuration, differential dust geometry, scattering of Ly$\alpha$ photons and the production efficiency of ionising photons compared to the UV luminosity, $\rm \xi_{\rm ion}$ \citep[see e.g.][]{Hayes2014,Dijkstra2017,MattheeGALEX2017,Shivaei2017}.

While a relation between $\rm f_{esc,Ly\alpha}$ and EW$_0$ is expected, we can investigate if it simply follows what would be predicted given that both the UV and Ly$\alpha$ trace SFRs. In order to predict $\rm f_{esc,Ly\alpha}$ based on Ly$\alpha$ EW$_0$ we first follow \citetalias{Dijkstra2010} who used the \cite{Kennicutt1998} SFR calibrations for a Salpeter IMF and UV continuum measured (observed) at 1400\,{\AA} to derive:
\begin{equation}\label{DW10_eq1}
\rm \frac{SFR_{Ly\alpha}}{SFR_{UV}}= f_{esc,Ly\alpha(DW10)}=\Big(\frac{C}{E}\Big)EW_{0},
\end{equation}
where $C=\Big(\frac{\nu_{\rm Ly\alpha}}{\nu_{\rm UV}}\Big)^{-2-\beta}\approx1.152^{-\beta-2}$ and $\beta$ is the UV slope (where L$_{\lambda}\propto\lambda^{\beta}$). The \cite{Kennicutt1998} SFR calibrations\footnote{Assuming a Ly$\alpha$/H$\alpha$ case B recombination coefficient of 8.7.} yield:
\begin{equation}
 \rm E=\frac{1.4\times10^{-28}\lambda_{Ly\alpha}}{\frac{7.9}{8.7}\times10^{-42}\nu_{Ly\alpha}} = 76.0\,{\rm \AA},
\end{equation}
which allows a final parameterisation of $\rm f_{esc,Ly\alpha}$ as a function of EW$_0$ and with just one free parameter, the UV $\beta$ slope:
\begin{equation}\label{final_DW10_s}
\rm f_{esc,Ly\alpha(DW10)}=\frac{1.152^{-\beta-2}}{76}EW_0
\end{equation}
The \citetalias{Dijkstra2010} methodology implicitly assumes a ``canonical", constant $\rm \xi_{\rm ion}=1.3\times10^{25}$\,Hz\,erg$^{-1}$ \citep{Kennicutt1998}\footnote{$\rm \xi_{ion} = 1.3\times10^{25} \frac{SFR_{\rm H\alpha}}{SFR_{UV}} \, \rm (Hz\,erg^{-1})$.}, and a unit ratio between Ly$\alpha$ and UV SFRs \citep[assuming 100\,Myr constant SFR; see also][and Equation \ref{This_study_DW10plus}]{Sobral2017_SC4K}.   \citetalias{Dijkstra2010} do not explicitly include the effect of dust in their framework which means assuming $0.0$ mag of extinction in the UV ($\rm A_{UV}=0.0$). Such framework will therefore typically overestimate the predicted $\rm f_{esc,Ly\alpha}$. Also, note that in \citetalias{Dijkstra2010} $\beta$ is simply a parameter used to extrapolate the UV continuum from rest-frame 1400\,{\AA} to 1216\,{\AA}, and thus no physical conditions change with $\beta$ \citep[but see e.g.][]{Popping2017,Narayanan2018}.

As in \citetalias{Dijkstra2010}, we use two different UV slopes: $\beta=-2.0$ and $\beta=-1.0$, which encompass the majority of LAEs\footnote{Note that a steeper $\beta$ (within the framework of \citetalias{Dijkstra2010}) results in an even more significant disagreement with observations for a fixed UV luminosity (measured at rest-frame 1400\,{\AA}; see \citetalias{Dijkstra2010}) or SFR, as $\beta$ is used to predict the UV continuum at $\approx1216$\,{\AA}. A steeper $\beta$ in this context leads to more UV continuum and a lower EW$_0$ for fixed SFR and $\rm f_{esc,Ly\alpha}$.} and result in $C=1.0$ and $C=0.87$, respectively ($C\approx1.152^{-\beta-2}$; see Equation \ref{final_DW10_s}). Based on the best empirical fits obtained in Section \ref{fits_empiri}, we would expect $C/E=0.0048$, which would yield $\beta\approx+5.13$. Indeed, allowing $\beta$ to vary freely within the \citetalias{Dijkstra2010} framework (Equation \ref{final_DW10_s}) allows to obtain relatively good fits to the data/observations (\,$\rm \chi^2_{reduced}\approx1.2$) but only for extremely red UV slopes of $\beta\approx+5$, which are completely excluded by other independent observations of LAEs. We therefore conclude that predicting $\rm f_{esc,Ly\alpha}$ based on the ratio of Ly$\alpha$ to UV SFRs using EW$_0$ and the \citetalias{Dijkstra2010} framework with realistic UV $\beta$ slopes significantly overestimates $\rm f_{esc,Ly\alpha}$ (as indicated by the dot-dashed lines in Figure \ref{EW_vs_f_esc}). Observations reveal higher Ly$\alpha$ EW$_0$ (by a factor of just over $\sim3$ higher than the canonical value) than expected for a given $\rm f_{esc,Ly\alpha}$. The results reveal processes that can boost the ratio between Ly$\alpha$ and UV (boosting EW$_0$), particularly by boosting Ly$\alpha$, or processes that reduce $\rm f_{esc,Ly\alpha}$.

%%%%%%%%%%%%%%%%%%%%%%%%%%%%%%%%%%%%%%%%%%%%%%%%
%
% FIGURE 2 - Relation between EW and Lya fesc - interpretation with E(B-V) and E_ion
%
%%%%%%%%%%%%%%%%%%%%%%%%%%%%%%%%%%%%%%%%%%%%%%%%
\begin{figure*}
\begin{tabular}{cc}
\includegraphics[width=8.88cm]{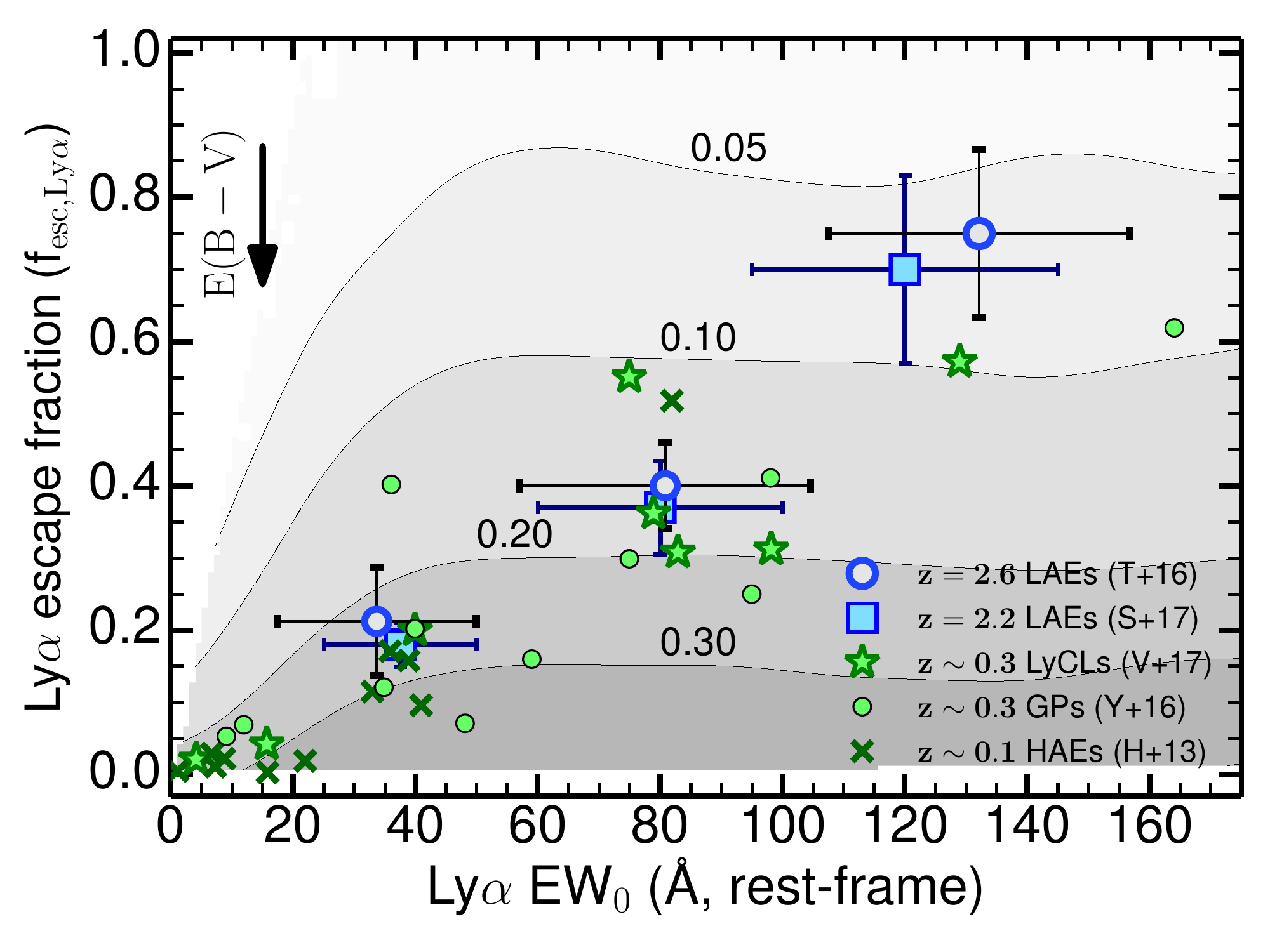}&  
\includegraphics[width=8.88cm]{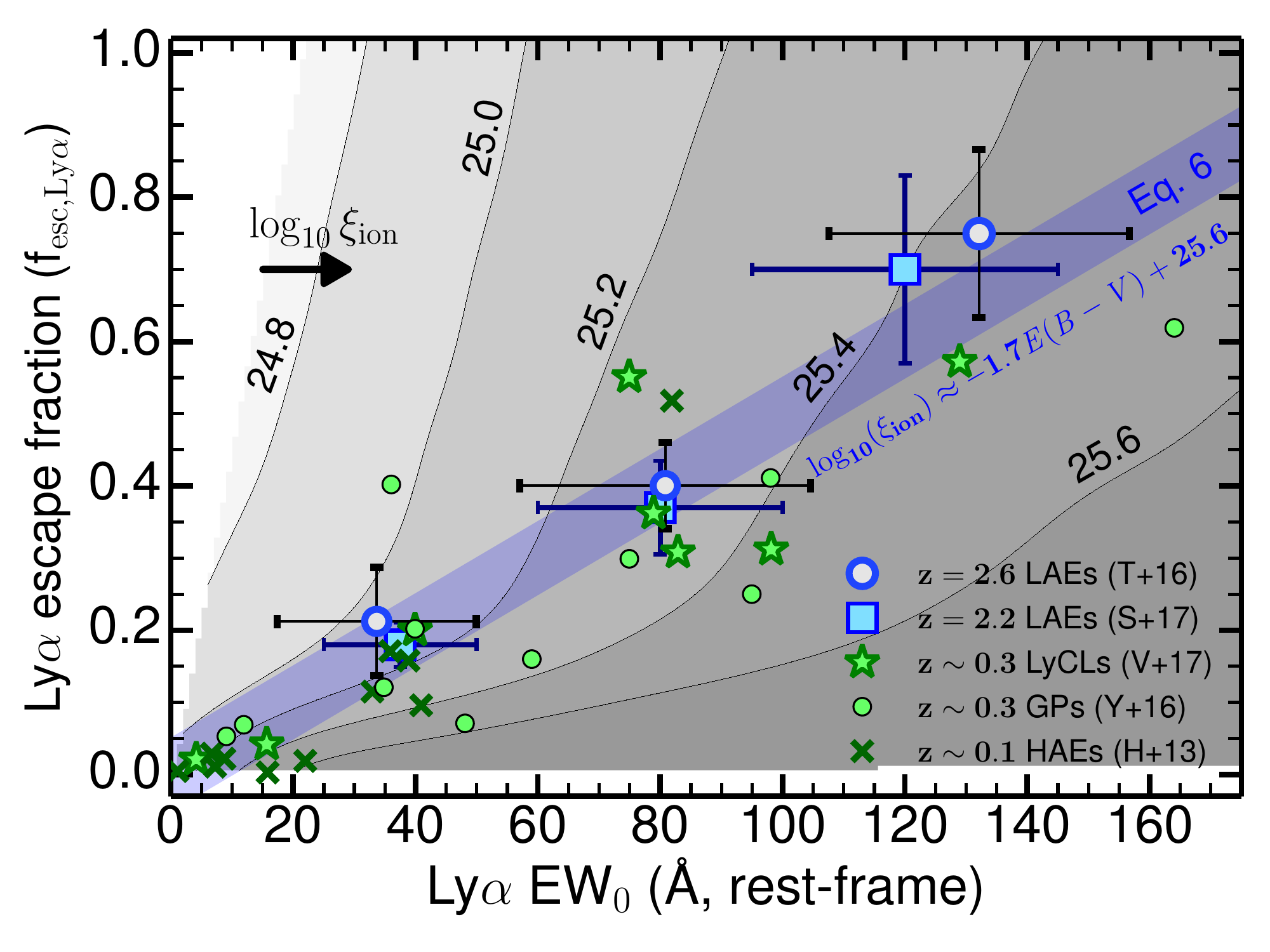}\\  
\end{tabular}
\caption{{\it Left:} The predicted $\rm f_{esc,Ly\alpha}$-Ly$\alpha$ EW$_0$ space for different $E(B-V)$ (contour levels) with our grid model (see \S\ref{phy_interp} and Appendix \ref{Toy_model}) and comparison with fits and implications by using Equation \ref{This_study_DW10plus} (right). We find that increasing dust extinction drives $\rm f_{esc,Ly\alpha}$ down for a fixed EW$_0$, with data at $z\sim0-2.6$ hinting for lower dust extinction at the highest EW$_0$ and higher dust extinction at the lowest EW$_0$, but with the range being relatively small overall and around $E(B-V)\approx0.05-0.3$. {\it Right:} The predicted $\rm f_{esc,Ly\alpha}$-EW$_0$ space for different $\rm \xi_{ion}$ (contours). We find that while increasing $E(B-V)$ mostly shifts the relation down, increasing $\rm \xi_{ion}$ moves the relation primarily to the right.}
\label{EW_vs_f_esc_interpret}
\end{figure*} 
%%%%%%%%%%%%%%%%%%%%%%%%%%%%

Potential explanations include scattering, (differential) dust extinction, excitation due to shocks originating from stellar winds and/or AGN activity, and short time-scale variations in SFRs, leading to a higher $\xi_{\rm ion}$ (see Figure \ref{EW_vs_f_esc}). High $\xi_{\rm ion}$ values ($\rm \xi_{\rm ion}\approx3\times10^{25}$\,Hz\,erg$^{-1}$) seem to be typical for LAEs \citep[e.g.][]{MattheeGALEX2017,Nakajima2018} and may explain the observed relation, but dust extinction likely also plays a role (see Figure \ref{EW_vs_f_esc} and Section \ref{phy_interp}). In order to further understand why the simple \citetalias{Dijkstra2010} framework fails to reproduce the observations (unless one invokes $\beta\approx+5$), we expand on the previous derivations by identifying the role of $\rm \xi_{\rm ion}$ (see derivations in \citealt{Sobral2017_SC4K}) and dust extinction ($\rm A_{UV}$) in setting the relation between Ly$\alpha$ and UV SFRs and thus we re-write the relation between $\rm f_{esc,Ly\alpha}$ and EW$_0$ as:
\begin{equation}\label{This_study_DW10plus}
\rm f_{esc,Ly\alpha} = \Big(\frac{1.152^{-\beta-2}}{76}EW_0\Big) \frac{1.3\times10^{25}}{\xi_{ion}}10^{-0.4A_{UV}}
\end{equation}
Note that Equation \ref{This_study_DW10plus} (this study) becomes Equation \ref{final_DW10_s} \citepalias{Dijkstra2010} if one assumes no dust extinction ($\rm A_{UV}=0$) and a canonical $\rm \xi_{\rm ion}=1.3\times10^{25}$\,Hz\,erg$^{-1}$. In order to keep the same framework as \citetalias{Dijkstra2010} and avoid spurious correlations and conclusions, here we also let $\beta$ be decoupled from $\rm A_{UV}$ \citep[but see][and Section \ref{phy_interp}]{Meurer1999}. By allowing all 3 parameters to vary ($\beta=[-2.4,-1.5]$, $\rm \log_{10}[\xi_{ion}/Hz\,erg^{-1}]=[24.5,26.5]$ and $\rm A_{UV}=[0,1]$) independently in order to attempt to fit observations, we find best fit values of $\beta=-2.0\pm0.3$, $\rm \log_{10}[\xi_{ion}/Hz\,erg^{-1}]=25.4\pm0.1$ and $\rm A_{UV}=0.5\pm0.3$ (corresponding to $E(B-V)=0.11\pm0.07$ with a \citealt{Calzetti00} dust law). We find that $\rm \log_{10}\xi_{ion}$ is the only parameter that is relatively well constrained within our framework and that there is a clear degeneracy/relation between $\rm \log_{10}\xi_{ion}$ and dust extinction (higher dust extinction allows for a lower $\rm \log_{10}\xi_{ion}$, with a relation given by ${\rm \log_{10}[\xi_{ion}/Hz\,erg^{-1}]\approx-1.71}E(B-V)+25.6$ with $\beta=-2.0\pm0.3$; see Figures \ref{EW_vs_f_esc_interpret} and \ref{Model_further_interp}) such that with no dust extinction one requires a high $\rm \log_{10}[\xi_{ion}/Hz\,erg^{-1}]=25.60\pm0.05$ while for $\rm A_{UV}\approx1.0$ a $\rm \log_{10}[\xi_{ion}/Hz\,erg^{-1}]\approx25.25$ is required to fit the observations (Figure \ref{Model_further_interp}). Observations of LAEs point towards $\rm \log_{10}[\xi_{ion}/Hz\,erg^{-1}]\approx25.5$ \citep[e.g.][]{MattheeGALEX2017,Nakajima2018}, in good agreement with our findings. If we fix $\rm \log_{10}[\xi_{ion}/Hz\,erg^{-1}]=25.5$, we still obtain a similar solution for $\beta$ (unconstrained), but we recover a lower $\rm A_{UV}=0.27\pm0.15$ (corresponding to $E(B-V)=0.06\pm0.04$ with a \citealt{Calzetti00} dust law), as we further break the degeneracy between $\rm A_{UV}$ and $\rm \xi_{ion}$. We find that canonical $\rm \log_{10}[\xi_{ion}/Hz\,erg^{-1}]=25.1$ values are strongly rejected and would only be able to explain the observations for significant amounts of dust extinction of $\rm A_{UV}\approx1.5-2.0$\,mag which are not found in typical LAEs.

In conclusion, we find that our modified analytical model (Equation \ref{This_study_DW10plus}, which expands the framework of \citetalias{Dijkstra2010}), is able to fit the observations relatively well. We find that high $\rm \xi_{ion}$ values of $\rm \log_{10}[\xi_{ion}/Hz\,erg^{-1}]=25.4\pm0.1$ and some low dust extinction ($E(B-V)\approx0.11$) are required to explain the observed relation between $\rm f_{esc,Ly\alpha}$ and EW$_0$. Without dust extinction one requires even higher ionisation efficiencies of $\rm \log_{10}[\xi_{ion}/Hz\,erg^{-1}]=25.60\pm0.05$. In general, the physical values required to explain observations agree very well with observations and further reveal that LAEs are a population with high $\rm \log_{10}[\xi_{ion}/Hz\,erg^{-1}]\approx25.4-25.6$ and low $E(B-V)\approx0.1$.

\subsection{The $\rm f_{esc,Ly\alpha}$-EW$_0$ relation: further physical interpretation}\label{phy_interp}

In order to further interpret the physics behind our observed empirical relation, we use a simple analytical toy model. In particular, we focus on the role of dust ($E(B-V)$) and $\xi_{\rm ion}$ (see details in Appendix \ref{Toy_model}). We independently vary SFRs, $E(B-V)$ and $\xi_{\rm ion}$ with flat priors to populate the $\rm f_{esc,Ly\alpha}$-EW$_0$ space. The toy model follows our framework using a \cite{Calzetti00} dust attenuation law and the \cite{Kennicutt1998} calibrations and relations between UV and H$\alpha$. We also assume the same nebular and stellar continuum attenuation \citep[see e.g.][]{Reddy2015} and use the \cite{Meurer1999} relation. We also vary some assumptions independently which depend on the binary fraction, stellar metallicity and the IMF, which include the intrinsic Ly$\alpha$/H$\alpha$ ratio, the intrinsic UV $\beta$ slope \citep[see e.g.][]{Wilkins2013} and $\rm f_{esc,LyC}$ (see e.g. Table \ref{Params_toy_model}). Furthermore, we introduce an extra parameter to further vary $\rm f_{esc,Ly\alpha}$ and mimic processes which are hard to model, such as scattering, which can significantly reduce or even boost $\rm f_{esc,Ly\alpha}$ \citep[][]{Neufeld1991} and allows our toy analytical model to sample a wide range of the $\rm f_{esc,Ly\alpha}$-EW$_0$ plane. We compute observed Ly$\alpha$ EW$_0$ and compare them with $\rm f_{esc,Ly\alpha}$ for 1,000,000 galaxy realisations. Further details are given in Appendix \ref{Toy_model}.

The key results from our toy model are shown in Figure \ref{EW_vs_f_esc_interpret}, smoothed with a Gaussian kernel of [0.07,20\,{\AA}] in the $\rm f_{esc,Ly\alpha}$-EW$_0$ parameter space. We find that both $E(B-V)$ and $\xi_{\rm ion}$ likely play a role in setting the $\rm f_{esc,Ly\alpha}$-EW$_0$ relation and changing it from simple predictions to the observed relation (see \S\ref{predict_from_UV}), a result which is in very good agreement with our findings in the previous section. As the left panel of Figure \ref{EW_vs_f_esc_interpret} shows, observed LAEs on the $\rm f_{esc,Ly\alpha}$-EW$_0$ relation seem to have low $E(B-V)\approx0.1-0.2$, with the lowest EW$_0$ sources displaying typically higher $E(B-V)$ of 0.2-0.3 and the highest EW$_0$ sources likely having lower $E(B-V)$ of $<0.1$. Furthermore, as the right panel of Figure \ref{EW_vs_f_esc_interpret} shows, high EW$_0$ LAEs have higher $\xi_{\rm ion}$, potentially varying from $\rm \log_{10}(\xi_{\rm ion}/Hz\,erg^{-1})\approx25$ to $\rm \log_{10}(\xi_{\rm ion}/Hz\,erg^{-1})\approx25.4$. Our toy model interpretation is consistent with recent results \citep[e.g.][]{Trainor2016,MattheeGALEX2017,Nakajima2018} for high EW$_0$ LAEs and with our conclusions in Section \ref{predict_from_UV}. Overall, a simple way to explain the $\rm f_{esc,Ly\alpha}$-EW$_0$ relation at $z\sim0-2.6$ is for LAEs to have narrow ranges of low $E(B-V)\approx0.1-0.2$, that may decrease slightly as a function of EW$_0$ and a relatively narrow range of high $\xi_{\rm ion}$ values that may increase with EW$_0$. Direct observations of Balmer decrements and of high excitation UV lines are required to confirm or refute our results.

Our toy model explores the full range of physical conditions independently without making any assumptions on how parameters may correlate, in order to interpret the observations in a simple unbiased way. However, the fact that observed LAEs follow a relatively tight relation between $\rm f_{esc,Ly\alpha}$ and EW$_0$ suggests that there are important correlations between e.g. dust, age and $\xi_{\rm ion}$. By selecting simulated sources in our toy model grid that lie on the observed relation (see Appendix \ref{Toy_model_interp}), we recover a tight correlation between $\xi_{\rm ion}$ and $E(B-V)$, while the full generated population in our toy model shows no correlation at all by definition (see Figure \ref{Model_further_interp}). This implies that the observed $\rm f_{esc,Ly\alpha}$-EW$_0$ relation could be a consequence of an evolutionary $\xi_{\rm ion}$-$E(B-V)$ sequence for LAEs, likely linked with the evolution of their stellar populations. For further details, see Appendix \ref{Toy_model_interp}. We note that the best fits to observations using Equation \ref{This_study_DW10plus} are consistent with this possible relation as the solutions follow a well defined anti-correlation between $\xi_{\rm ion}$ and dust extinction with a similar relation and slope; see Figure \ref{Model_further_interp} for a direct comparison.

\subsection{Estimating $f_{esc,Ly\alpha}$ with a simple observable: Ly$\alpha$ EW$_0$}

We find that LAEs follow a simple relation between $\rm f_{esc,Ly\alpha}$ and Ly$\alpha$ EW$_{0}$ roughly independently of redshift (for $z\leq2.6$). Motivated by this, we propose the following empirical estimator (see Table \ref{Fits_fesc_EW}) for $\rm f_{esc,Ly\alpha}$ as a function of Ly$\alpha$ EW$_{0}$ (\AA):
\begin{equation}\label{EW_fesc}
\rm f_{esc,Ly\alpha}=0.0048^{+0.0007}_{-0.0007}\,{\rm EW_{0}}\pm0.05 \ \ [\, {\rm 0\leq EW_0\leq160\,{\AA}}].
\end{equation}
This relation may hold up to EW$_{0}\approx210$\,{\AA}, above which we would predict $\rm f_{esc,Ly\alpha}\approx1$. This relation suggests that it is possible to estimate $\rm f_{esc,Ly\alpha}$ for LAEs within a scatter of 0.2\,dex even if only the Ly$\alpha$ EW$_{0}$ is known/constrained. It also implies that the observed Ly$\alpha$ luminosities are essentially equal to intrinsic Ly$\alpha$ luminosities for sources with EW$_{0}$ as high as $\approx$ 200 {\AA}. We conclude that while the escape of Ly$\alpha$ photons can depend on a range of properties in a very complex way \citep[see e.g.][]{Hayes2010,Matthee2016,Yang2017}, using EW$_0$ and Equation \ref{EW_fesc} leads to predicting $\rm f_{esc,Ly\alpha}$ within $\approx0.1-0.2$\,dex of real values. This compares with a larger scatter of $\approx0.3$\,dex for relations with derivative or more difficult quantities to measure such as dust extinction or the red peak velocity of the Ly$\alpha$ line \citep[e.g.][]{Yang2017}. We propose a linear relation for its simplicity and because current data do not suggest a more complex relation. Larger data-sets with H$\alpha$ and Ly$\alpha$ measurements, particularly those covering a wider parameter space (e.g. different sample selections, multiple redshifts and both high and low EWs), may lead to the necessity of a more complicated functional form. A departure from a linear fit may also provide further insight of different physical processes driving the relation and the scatter (e.g. winds, orientation angle, burstiness or additional ionisation processes such as fluorescence).

Equation \ref{EW_fesc} may thus be applied to estimate $\rm f_{esc,Ly\alpha}$ for a range of LAEs in the low and higher redshift Universe. For example, the green pea J1154+2443 \citep{Izotov2017_Lya}, has a measured $\rm f_{esc,Ly\alpha}$ directly from dust corrected H$\alpha$ luminosity  of $\approx0.7-0.8$\footnote{This may be up to $\approx0.98$ if H$\beta$ is used; see \citep{Izotov2017_Lya}.}, while Equation \ref{EW_fesc} would imply $\approx0.6-0.7$ based on the $EW_0\approx133$\,{\AA} for Ly$\alpha$, thus implying a difference of only 0.06-0.1\,dex. Furthermore, in principle, Equation \ref{EW_fesc} could also be explored to transform EW$_0$ distributions \citep[e.g.][and references therein]{Hashimoto2017} into distributions of $\rm f_{esc,Ly\alpha}$ for LAEs.

%%%%%%%%%%%%%%%%%%%%%%%%%%%%%%%%%%%%
%
% FIGURE 3 - Left: how well one can derive SFR(Ha) for sources we can measure it directly
%
%%%%%%%%%%%%%%%%%%%%%%%%%%%%%%%%%%%%
\begin{figure*}
\centering
\includegraphics[width=13.5cm]{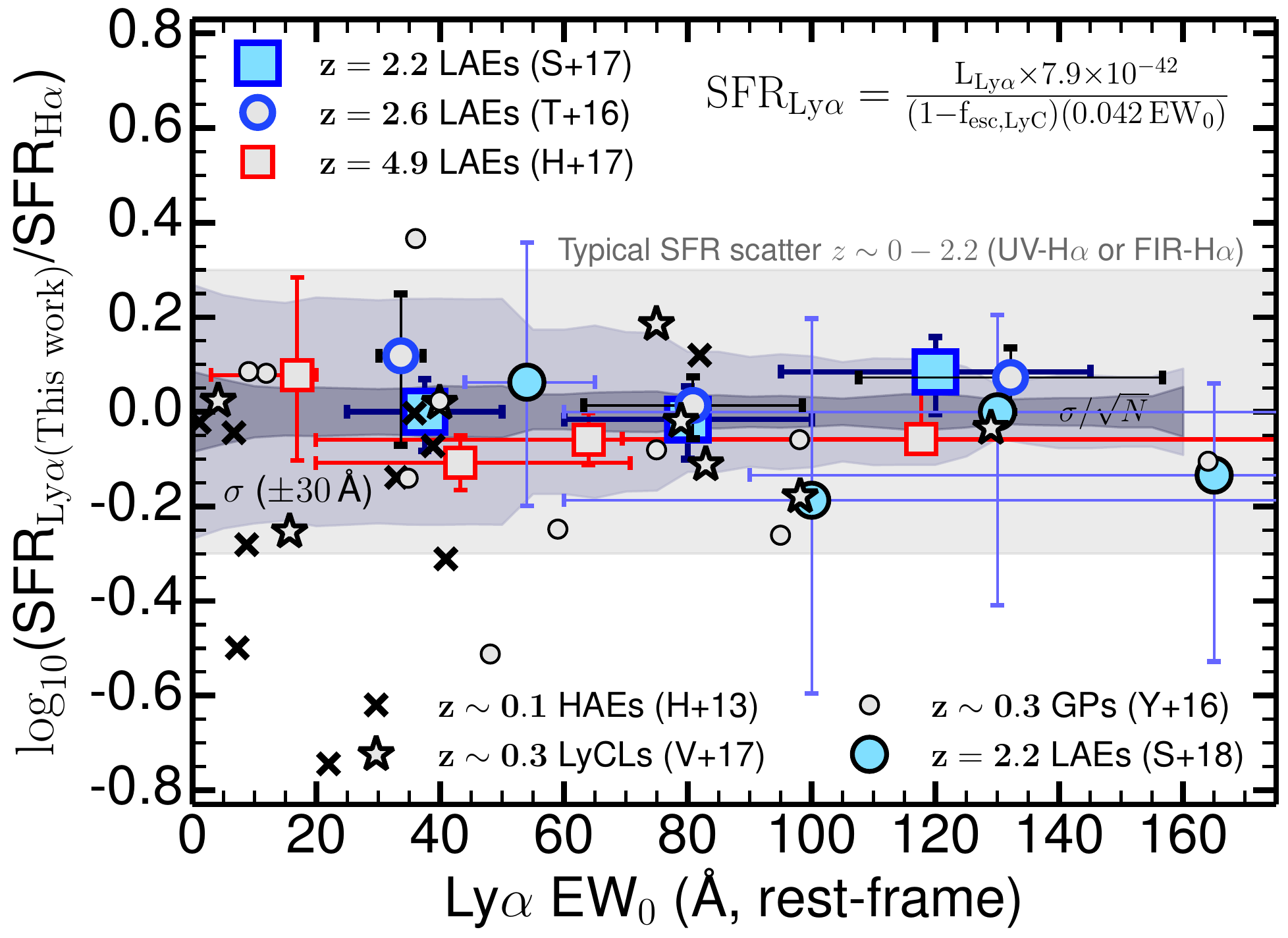}  
\caption{The logarithmic ratio between SFRs computed with Equation \ref{Lya_SFR_this} using Ly$\alpha$ luminosity and EW$_0$ and the ``true" SFR, measured directly from dust-corrected H$\alpha$ luminosity (given our definitions, $\log_{10}$(SFR$_{\rm Ly\alpha(This\,\,work)}$/SFR$_{\rm H\alpha})=\rm \log_{10}(f_{\rm esc,Ly\alpha(H\alpha)}/f_{\rm esc,Ly\alpha(This\,\,work)})$). The standard deviation ($\sigma$) and $\sigma$/$\sqrt{N}$ (N: number of sources) are computed in $\pm30$\,{\AA} bins and indicated in light and darker grey. We find a relatively small scatter which may decrease for higher EWs and that is at the global level of $\pm0.12$\,dex for the typical definition of LAE at higher redshift (EW$_0>20$\,{\AA}), but rises to $\approx0.2$\,dex at the lowest EWs. We also provide a comparison of the typical scatter between UV and FIR SFRs in relation to H$\alpha$ at $z\sim0-2$ \citep[$\approx0.3$\,dex; see e.g.][]{Dom2012,Oteo2015}.}
\label{Residuals_relation}
\end{figure*} 
%%%%%%%%%%%%%%%%%%%%%%%%%%%%

\subsection{Ly$\alpha$ as an SFR indicator: empirical calibration and errors}\label{emp_SFR_err}

Driven by the simple relation (Equation \ref{EW_fesc}) found up to $z\sim2.6$, we derive an empirical calibration to obtain SFRs based on two simple, direct observables for LAEs at high redshift: 1) Ly$\alpha$ EW$_{0}$ and 2) observed Ly$\alpha$ luminosity. This calibration is based on observables, but predicts the dust-corrected SFR\footnote{We use extinction corrected H$\alpha$ luminosities.}. Based on Equations \ref{eq:SFR} and \ref{EW_fesc}, for a Salpeter (Chabrier) IMF we can derive\footnote{Note that the constant 0.042 has units of {\AA}$^{-1}$, and results from $8.7\times0.0048$\,{\AA}$^{-1}$. Also, note that the relation is valid for $\rm 0\leq EW_0\leq160$\,{\AA} following Equation \ref{EW_fesc}. For EW$_0>160$\,{\AA} the relation has not been calibrated yet. Furthermore, if the relation is to be used at even higher EWs, then for EW$_0>207$\,{\AA} the factor $0.042\,{\rm EW}_{0}$ should be set to 8.7 (or the appropriate/assumed case B recombination constant), corresponding to a $\approx100$\,\% escape fraction of Ly$\alpha$ photons.}:
\begin{equation}\label{Lya_SFR_this}
{\rm SFR_{Ly\alpha}\,[{\rm M_{\odot}\,yr^{-1}}]}=\frac{{\rm L}_{\rm Ly\alpha}\times7.9\,(4.4)\times10^{-42}}{\rm (1-f_{esc,LyC})(0.042\,{\rm EW}_{0})} \, (\pm15\%)
\end{equation}

The current best estimate of the scatter in Equation \ref{EW_fesc} (the uncertainty in the relation to calculate $\rm f_{esc,Ly\alpha}$ is $\pm0.05$) implies a $\pm0.07$\,dex uncertainty in the extinction corrected SFRs from Ly$\alpha$ with our empirical calculation. In order to investigate other systematic errors, we conduct a Monte Carlo analysis by randomly varying $\rm f_{esc,LyC}$ (0.0 to 0.2) and the case B coefficient (from 8.0 to 9.0), along with perturbing $\rm f_{esc,Ly\alpha}$ from $-0.05$ to $+0.05$. We assume that all properties are independent, and thus this can be seen as a conservative approach to estimate the uncertainties. We find that the uncertainty in $\rm f_{esc,Ly\alpha}$ is the dominant source of uncertainty (12\%) with the uncertainty on $\rm f_{esc,LyC}$ and the case B coefficient contributing an additional 3\% for a total of 15\%. This leads to an expected uncertainty of Equation \ref{Lya_SFR_this} of 0.08\,dex.

Note that the SFR calibration presented in equation \ref{Lya_SFR_this} follows \citet{Kennicutt1998} and thus a solar metallicity, which may not be be fully applicable to LAEs, typically found to be sub-solar \citep[][]{NakajimaOuchi2013,Steidel2016,Suzuki2017,Sobral2018b}. Other caveats include the applicability of the \citet{Calzetti00} dust law \citep[see e.g.][]{Reddy2016} and the shape and slope of the IMF used, although any other SFR calibration/estimator will share similar caveats.

\subsection{Ly$\alpha$ as an SFR indicator: performance and implications}

In Figure \ref{Residuals_relation} we apply Equation \ref{Lya_SFR_this} to compare the estimated SFRs (from Ly$\alpha$) with those computed with dust corrected H$\alpha$ luminosities. We also include individual sources at $z\sim2.2$ \citep[S18;][]{Sobral2018b} and recent results from \cite{Harikane2017} at $z=4.8$ which were not used in the calibration, and thus provide an independent way to test our new calibration. We find a global scatter of $\approx0.12$\,dex, being apparently larger for lower EW$_0$, but still lower than the typical scatter between SFR indicators after dust corrections (e.g. UV-H$\alpha$ or FIR-H$\alpha$; see \citealt{Dom2012,Oteo2015}), as shown in Figure \ref{Residuals_relation}. The small scatter and approximately null offset between our calibration's prediction and measurements presented by \cite{Harikane2017} at $z\sim5$ suggest that Equation \ref{Lya_SFR_this} may be applicable at higher redshift with similarly competitive uncertainties (see \S\ref{highz_appli_LAEs} and \S\ref{highz_appli}). Nonetheless, we note that the measurements presented by \cite{Harikane2017} are inferred from broad-band IRAC photometry/colours as it is currently not possible to directly measure H$\alpha$ line luminosities beyond $z\sim2.5$, and thus any similar measurements should be interpreted with some caution.

\subsection{Application to bright and faint LAEs at high redshift}\label{highz_appli_LAEs}

Our new empirical calibration of Ly$\alpha$ as an SFR indicator allows to estimate SFRs of LAEs at high redshift. The global Ly$\alpha$ luminosity function at $z\sim3-6$ has a typical Ly$\alpha$ luminosity (L$^{\star}_{\rm Ly\alpha}$) of 10$^{42.9}$\,erg\,s$^{-1}$ \citep[see e.g.][and references therein]{Drake2017b,Herenz2017,Sobral2017_SC4K}, with these LAEs having EW$_0\approx80$\,{\AA} (suggesting $\rm f_{esc,Ly\alpha}=0.38\pm0.05$ with Equation \ref{EW_fesc}), which implies SFRs of $\approx20$\,M$_{\odot}$\,yr$^{-1}$. If we explore the public SC4K sample of LAEs at $z\sim2-6$ \citep[][]{Sobral2017_SC4K}, limiting it to sources with up to EW$_0=210$\,{\AA} and that are consistent with being star-forming galaxies \citep[$\rm L_{Ly\alpha}<10^{43.2}$\,erg\,s$^{-1}$; see][]{Sobral2018b}, we find a median SFR for LAEs of $12^{+9}_{-5}$\,M$_{\odot}$\,yr$^{-1}$, ranging from $\approx2$\,M$_{\odot}$\,yr$^{-1}$ to $\approx90$\,M$_{\odot}$\,yr$^{-1}$ at $z\sim2-6$. These reveal that ``typical" to luminous LAEs are forming stars below and up to the typical SFR (SFR$^{\star}\approx40-100$\,M$_{\odot}$\,yr$^{-1}$) at high redshift \citep[see][]{Smit2012,Sobral2014}; see also \cite{Kusakabe2018}.

Deep MUSE Ly$\alpha$ surveys \citep[e.g.][]{Drake2017a,Hashimoto2017} are able to sample the faintest LAEs with a median L$_{\rm Ly\alpha}=10^{41.9\pm0.1}$\,erg\,s$^{-1}$ and EW$_0=87\pm6$ \citep[][]{Hashimoto2017} at $z\sim3.6$. We predict a typical $\rm f_{esc,Ly\alpha}=0.42\pm0.05$ and SFR$_{\rm Ly\alpha}=1.7\pm0.3$\,M$_{\odot}$\,yr$^{-1}$ for those MUSE LAEs. Furthermore, the faintest LAEs found with MUSE have L$_{\rm Ly\alpha}=10^{41}$\,erg\,s$^{-1}$ \citep[][]{Hashimoto2017}, implying SFRs of $\approx0.1$\,M$_{\odot}$\,yr$^{-1}$ with our calibration. Follow-up {\it JWST} observations targeting the H$\alpha$ line for faint MUSE LAEs are thus expected to find typical H$\alpha$ luminosities of $2\times10^{41}$\,erg\,s$^{-1}$ and as low as $\approx1-2\times10^{40}$\,erg\,s$^{-1}$ for the faintest LAEs. Based on our predicted SFRs, we expect MUSE LAEs to have UV luminosities from $\rm M_{UV}\approx-15.5$ for the faintest sources \cite[see e.g.][]{Maseda2018}, to $\rm M_{UV}\approx-19$ for more typical LAEs, thus potentially linking faint LAEs discovered from the ground with the population of SFGs that dominate the faint end of the UV luminosity function \citep[e.g.][]{Fynbo2003,GronkeDi2015,Dressler2015}.

\subsection{Comparison with UV and implications at higher redshift}\label{highz_appli}

Equations \ref{EW_fesc} and \ref{Lya_SFR_this} can be applied to a range of spectroscopically confirmed LAEs in the literature. We also extend our predictions to sources within the epoch of re-ionisation, although there are important caveats on how the Ly$\alpha$ transmission is affected by the IGM; see e.g. \cite{Laursen2011}.

We explore a recent extensive compilation by \cite{Matthee2017followup} of both Ly$\alpha$- and UV-selected LAEs with spectroscopic confirmation and Ly$\alpha$ measurements \citep[e.g.][]{Ouchi2008,Ouchi2009,Ono2012,Sobral2015,Zabl2015,Stark2015,Ding2017,Shibuya2017_spec}; see Table \ref{HIghz_appl_tabl}. These include published L$_{\rm Ly\alpha}$, EW$_0$ and M$_{\rm UV}$. In order to correct UV luminosities we use the UV $\beta$ slope, typically used to estimate $\rm A_{UV}$\footnote{We use $A_{\rm UV}=4.43+1.99\beta$; see \cite{Meurer1999}, but see also discussions on uncertainties and limitations in e.g. \cite{Popping2017}, \cite{Narayanan2018} and references therein.}. We use $\beta$ values (and errors) estimated in the literature for each source when available. When individual $\beta$ values are not available, we use $\beta=-1.6\pm0.2$ for UV-selected sources \citep[typical for their UV luminosity; e.g.][]{Bouwens2009}, while for the luminous LAEs we use $\beta=-1.9\pm0.2$. As a comparison, we also use a fixed $\beta=-1.6\pm0.2$ for all the sources, which leads to a correction of $\rm A_{UV}\approx1.25$\,mag. We list UV $\beta$ slopes and resulting SFRs in Table \ref{HIghz_appl_tabl}.

\begin{figure}
\centering
\includegraphics[width=8.8cm]{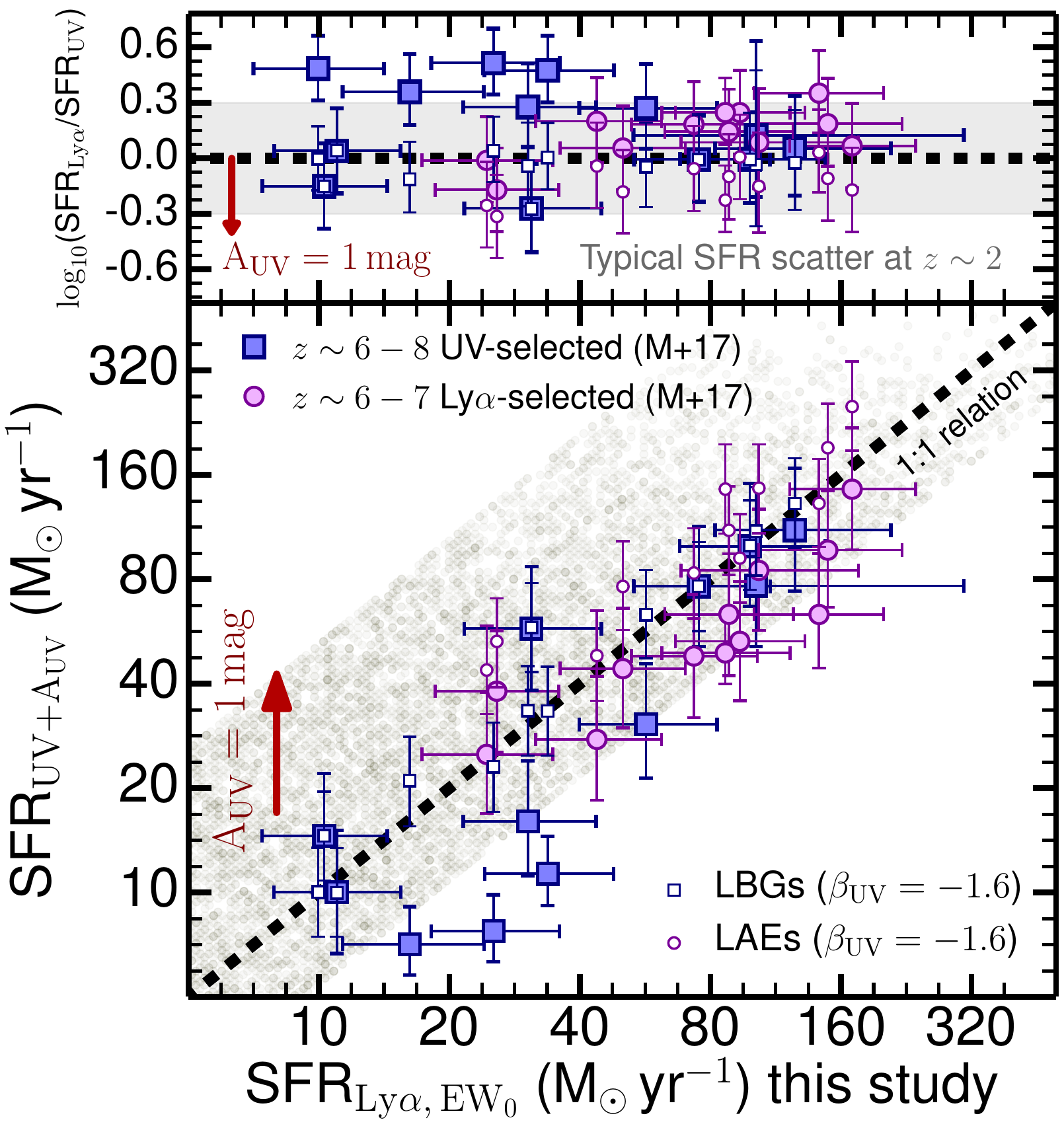}  
\caption{Comparison between SFRs computed with our new empirical calibration for Ly$\alpha$ as an SFR indicator (Equation \ref{Lya_SFR_this}) and those computed based on dust corrected UV luminosity (see \S\ref{highz_appli}) for a compilation of $z\sim5-8$ sources \citep[see][and references therein]{Matthee2017followup}. Our simple empirical calibration of Ly$\alpha$ as a SFR is able to recover dust corrected UV SFRs for the most star-forming sources, with a typical scatter of $\approx0.2-0.3 $\,dex (the scatter is lower if one assumes a fixed $\beta=-1.6\pm0.2$). For the sources with the lowest UV SFRs we find that Equation \ref{Lya_SFR_this} seems to over-predict SFRs, potentially due to IGM effects which can lead to a lower EW$_0$. We also compute SFRs in the same way with observables from our toy model and show the results of all realisations in grey. We find that the scatter in our toy model is much larger, with this being driven by $E(B-V)$ being able to vary from 0.0 to 0.5.}
\label{SFR_vs_SFR}
\end{figure}

We predict (dust-corrected) SFRs using L$_{\rm Ly\alpha}$ and EW$_0$ only (Equation \ref{Lya_SFR_this}) and compare with SFRs measured from dust-corrected UV luminosities \citep[][]{Kennicutt1998}; see Table \ref{HIghz_appl_tabl}. We make the same assumptions and follow the same methodology to transform the observables of our toy model/grid into SFRs (see Figure \ref{SFR_vs_SFR}). We note that, as our simulation shows, one expects a correlation even if our calibration of Ly$\alpha$ as an SFR indicator is invalid at high redshift, but our grid shows that the scatter depends significantly on dust extinction. Therefore, we focus our discussion on the normalisation of the relation and particularly on the scatter, not on the existence of a relation. We also note that our calibration is based on dust corrected H$\alpha$ luminosities at $z\sim0-2.6$, and that UV luminosities are not used prior to this Section.

%%%%%%%%%%%%%%%%%%%%%%%%%%%%%%%%%%%%
%
% FIGURE 4 - Comparison SFR inferred vs measured from other means
%
%%%%%%%%%%%%%%%%%%%%%%%%%%%%%%%%%%%%
%%%%%%%%%%%%%%%%%%%%%%%%%%%%

Our results are shown in Figure \ref{SFR_vs_SFR} (see Table \ref{HIghz_appl_tabl} for details on individual sources), which contains sources at a variety of redshifts, from $z\sim6$ to $z\sim8$ \citep[e.g.][]{Oesch2015,Stark2017}. We find a good agreement between our predicted Ly$\alpha$ SFRs based solely on Ly$\alpha$ luminosities and EW$_0$ and the dust corrected UV SFRs for sources with the highest SFRs at $z\sim6-8$ (Figure \ref{SFR_vs_SFR}), with a scatter of $\approx0.2-0.3$\,dex. Interestingly, Equation \ref{Lya_SFR_this} seems to over-predict (compared to the UV) Ly$\alpha$ SFRs for the least star-forming sources ($\leq30-40$\,M$_{\odot}$\,yr$^{-1}$). This is caused by their typically very low EW$_0$, which would imply a low $\rm f_{esc,Ly\alpha}$, thus boosting the Ly$\alpha$ SFR compared to the UV. Taken as a single population, the UV-selected sources (LBGs) show a higher $\rm \log_{10}(SFR_{Ly\alpha}/SFR_{UV})=0.23\pm0.24$ than LAEs that reveal $\rm \log_{10}(SFR_{Ly\alpha}/SFR_{UV})=0.15\pm0.13$. Such discrepancies could be caused by the IGM which could be reducing the EW$_0$ and $\rm f_{esc,Ly\alpha}$. This would happen preferentially for the UV selected and for the sources with the lowest SFRs without strong Ly$\alpha$ in a way that our calibration at $z\sim0-2.6$ simply does not capture. However, the deviation from a ratio of 1 is not statistically significant given the uncertainties and there is a large scatter from source to source to be able to further quantify the potential IGM effect.

Overall, our results and application to higher redshift reveals that Equation \ref{Lya_SFR_this} is able to retrieve SFRs with very simple observables even for LAEs within re-ionisation \citep[e.g.][]{Ono2012,Stark2015,Stark2017,Schmidt2017}, provided they are luminous enough. In the early Universe the fraction of sources that are LAEs is higher \citep[e.g.][]{Stark2010,Stark2017,Caruana2018}, thus making our calibration potentially applicable to a larger fraction of the galaxy population, perhaps with an even smaller scatter due to the expected narrower range of physical properties and more compact sizes \citep[see discussions in e.g.][]{Paulino-Afonso2017}. Our calibration of Ly$\alpha$ as an SFR indicator is simple, directly calibrated with H$\alpha$, and should not have a significant dependence on metallicity, unlike other proposed SFRs tracers at high redshift such as [C{\sc ii}] luminosity or other weak UV metal lines.

It is surprising that our calibration apparently still works even at $z\sim7-8$ for the most luminous LAEs. This seems to indicate that the IGM may not play a significant role for these luminous Ly$\alpha$-visible sources, potentially due to early ionised bubbles \citep[see e.g.][]{Matthee2015,MattheeCOLA1,Mason2018a,Mason2018b} or velocity offsets of Ly$\alpha$ with respect to systemic \citep[see e.g.][]{Stark2017}. Interestingly, we find offsets between our calibration and the computed UV SFRs for the faintest sources, hinting that IGM effects start to be much more noticeable for such faint sources which may reside in a more neutral medium and/or on smaller ionised bubbles. Further observations measuring the velocity offsets between the Ly$\alpha$ and systemic redshifts for samples of LAEs within the epoch of re-ionisation and those at $z\sim3-5$ will allow to check and test the validity of the relation within the epoch of re-ionisation.

\subsection{A tool for re-ionisation: predicting the LyC luminosity}\label{highz_LyC}

Based on our results and assumptions (see \S\ref{Measuring_fesc}), we follow \cite{MattheeGALEX2017}\footnote{We assume $\rm f_{dust}\approx0$ \citep[see][]{MattheeGALEX2017}, i.e., we make the assumption that for LAEs the dust extinction to LyC photons within HII regions is $\approx0$.} and derive a simple expression to predict the number of produced LyC photons per second, $\rm Q_{ion}$ (s$^{-1}$) with direct Ly$\alpha$ observables (L$_{\rm Ly\alpha}$ and EW$_0$)\footnote{Note that the relation is valid for $\rm 0\leq EW_0\leq160$\,{\AA} following Equation \ref{EW_fesc}. If the relation is to be used beyond the calibrated range then for EW$_0>207$\,{\AA} the factor $0.042\,{\rm EW}_{0}$ should be set to 8.7 (case B recombination), corresponding to a $\approx100$\,\% escape fraction of Ly$\alpha$ photons. Note that one can also vary the Ly$\alpha$/H$\alpha$ case B ratio between 8.0 and 9.0 to better sample systematic errors due to unknown gas temperatures, although these result in small systematic variations.}:
\begin{equation}\label{LyC_preditor}
\rm Q_{ion,Ly\alpha}\,[{\rm s^{-1}}]=\frac{{\rm L}_{\rm Ly\alpha}}{\rm c_{H\alpha}\,(1-f_{esc,LyC})\,(0.042\,{\rm EW}_{0})},
\end{equation} 
where $\rm c_{H\alpha}=1.36\times10^{-12}$\,erg \citep[e.g.][]{Kennicutt1998,Schaerer2003}, under our case B recombination assumption (see \S\ref{Measuring_fesc}). We caution that Equation \ref{LyC_preditor} may not be fully valid for all observed LAEs within the epoch of reionisation. This is due to possible systematic effects on EW$_0$ of an IGM which is partly neutral, although we note that as found in Section \ref{highz_appli} it may well be valid for the most luminous LAEs at $z\sim7-8$.

Recent work by e.g. \cite{Verhamme2017} show that LyC leakers are strong LAEs, and that $\rm f_{esc,Ly\alpha}$ is linked and/or can be used to predict $\rm f_{esc,LyC}$ \citep[see][]{Chisholm2018}. Equation \ref{LyC_preditor} provides an extra useful tool: an empirical simple estimator of $\rm Q_{ion}$ for LAEs given observed Ly$\alpha$ luminosities and EW$_0$. Note that Equation \ref{LyC_preditor} does not require measuring UV luminosities or $\xi_{\rm ion}$, but instead direct, simple observables. \cite{Matthee2017followup} already used a similar method to predict $\xi_{\rm ion}$ at high redshift. Coupled with an accurate estimate of the escape fraction of LyC photons from LAEs \citep[see e.g.][]{Steidel2018}, a robust estimate of the full number density of LAEs from faint to the brightest sources \citep[][]{Sobral2017_SC4K} and their redshift evolution, Equation \ref{LyC_preditor} may provide a simple tool to further understand if LAEs were able to re-ionise the Universe.

\section{Conclusions}\label{conclusions}

Ly$\alpha$ is intrinsically the brightest emission-line in active galaxies, and should be a good SFR indicator. However, the uncertain and difficult to measure $\rm f_{esc,Ly\alpha}$ has limited the interpretation and use of Ly$\alpha$ luminosities. In order to make progress, we have explored samples of LAEs at $z=0-2.6$ with direct Ly$\alpha$ escape fractions measured from dust corrected H$\alpha$ luminosities which do not require any SED fitting, $\rm \xi_{ion}$ or other complex assumptions based on derivative quantities. Our main results are:

\begin{description}

\item[$\bullet$\,\,] There is a simple, linear relation between $\rm f_{esc,Ly\alpha}$ and Ly$\alpha$ EW$_0$: $\rm f_{esc,Ly\alpha}=0.0048\,EW_0[{\AA}]\pm0.05$ (Equation \ref{EW_fesc}) which is shallower than simple expectations, due to both more ionising photons per UV luminosity ($\rm \xi_{ion}$) and low dust extinction ($E(B-V)$) for LAEs (Figure \ref{EW_vs_f_esc}). This allows the prediction of $\rm f_{esc,Ly\alpha}$ based on a simple direct observable, and thus to compute the intrinsic Ly$\alpha$ luminosity of LAEs at high redshift.

\smallskip

\item[$\bullet$\,\,]  The observed $\rm f_{esc,Ly\alpha}$-EW$_0$ can be explained by high $\rm \xi_{ion}$ and low $E(B-V)$ or, more generally, by a tight $\rm \xi_{ion}$-$E(B-V)$ sequence for LAEs, with higher $\rm \xi_{ion}$ implying lower $E(B-V)$ and vice versa. $\rm \xi_{ion}$ and $E(B-V)$ may vary within the $\rm f_{esc,Ly\alpha}$-EW$_0$ plane (Figure \ref{EW_vs_f_esc_interpret}). Our results imply that the higher the EW$_0$ selection, the higher the $\rm \xi_{ion}$ and the lower the $E(B-V)$.

\smallskip

\item[$\bullet$\,\,] The $\rm f_{esc,Ly\alpha}$-EW$_0$ relation reveals a scatter of only 0.1-0.2\,dex for LAEs, and there is evidence for the relation to hold up to $z\sim5$ (Figure \ref{Residuals_relation}). The scatter is higher towards lower EW$_0$, consistent with a larger range in galaxy properties for sources with the lowest EW$_0$. At the highest EW$_0$, on the contrary, the scatter may be as small as $\approx0.1$\,dex, consistent with high EW$_0$ LAEs being an even more homogeneous population of dust-poor, high ionisation star-forming galaxies.

\smallskip

\item[$\bullet$\,\,] We use our results to calibrate Ly$\alpha$ as an SFR indicator for LAEs (Equation \ref{Lya_SFR_this}) and find a global scatter of 0.2\,dex between measurements using Ly$\alpha$ only and those using dust-corrected H$\alpha$ luminosities. Such scatter seems to depend on EW$_0$, being larger at the lowest EW$_0$. Our results also allow us to derive a simple estimator of the number of LyC photons produced per second (Equation \ref{LyC_preditor}) with applications to studies of the epoch of re-ionisation.

\smallskip

\item[$\bullet$\,\,] Equation \ref{Lya_SFR_this} implies that star-forming LAEs at $z\sim2-6$ have SFRs typically ranging from 0.1 to 20\,M$_{\odot}$\,yr$^{-1}$, with MUSE LAEs expected to have typical SFRs of $1.7\pm0.3$\,M$_{\odot}$\,yr$^{-1}$, and more luminous LAEs having SFRs of $12^{+9}_{-5}$\,M$_{\odot}$\,yr$^{-1}$. 

\smallskip

\item[$\bullet$\,\,] SFRs based on Equation \ref{Lya_SFR_this} are in good agreement with dust corrected UV SFRs even within the epoch of re-ionisation for SFRs higher than $\approx30-40$\,M$_{\odot}$\,yr$^{-1}$, hinting for it to be applicable in the very early Universe for bright enough LAEs. For lower SFRs we find that Equation \ref{Lya_SFR_this} may over-predict SFRs compared to the UV, potentially due to IGM effects. If shown to be the case, our results have implications for the minor role of the IGM in significantly changing Ly$\alpha$ luminosities and EW$_0$ for the most luminous LAEs within the epoch of re-ionisation.

\end{description}

\noindent Our results provide a simple interpretation of the tight $\rm f_{esc,Ly\alpha}$-EW$_0$ relation. Most importantly, we provide simple and practical tools to estimate $\rm f_{esc,Ly\alpha}$ at high redshift with two direct observables and thus to use Ly$\alpha$ as an SFR indicator and to measure the number of ionising photons from LAEs. The empirical calibrations presented here can be easily tested with future observations with {\it JWST} which can obtain H$\alpha$ and H$\beta$ measurements for high-redshift LAEs.

\begin{acknowledgements}

We thank the anonymous referees for multiple comments and suggestions which have improved the manuscript. JM acknowledges the support of a Huygens PhD fellowship from Leiden University. We have benefited greatly from the publicly available programming language {\sc Python}, including the {\sc NumPy} \& {\sc SciPy} \citep[][]{van2011numpy,jones}, {\sc Matplotlib} \citep[][]{Hunter:2007} and {\sc Astropy} \citep[][]{Astropy2013} packages, and the {\sc Topcat} analysis program \citep{Topcat}. The results and samples of LAEs used for this paper are publicly available \citep[see e.g.][]{Sobral2017,Sobral2017_SC4K} and we also provide the toy model used as a {\sc python} script.

\end{acknowledgements}

\bibliography{bib_LAEevo.bib}

\begin{thebibliography}{116}
\expandafter\ifx\csname natexlab\endcsname\relax\def\natexlab#1{#1}\fi

\bibitem[{{An} {et~al.}(2017){An}, {Zheng}, {Hao}, {Huang}, \& {Xia}}]{An2017}
{An}, F.~X., {Zheng}, X.~Z., {Hao}, C.-N., {Huang}, J.-S., \& {Xia}, X.-Y.
  2017, \apj, 835, 116

\bibitem[{{Astropy Collaboration} {et~al.}(2013)}]{Astropy2013}
{Astropy Collaboration} {et~al.} 2013, A\&A, 558, A33

\bibitem[{{Atek} {et~al.}(2008){Atek}, {Kunth}, {Hayes}, {{\"O}stlin}, \&
  {Mas-Hesse}}]{Atek2008}
{Atek}, H., {Kunth}, D., {Hayes}, M., {{\"O}stlin}, G., \& {Mas-Hesse}, J.~M.
  2008, \aap, 488, 491

\bibitem[{{Bacon} {et~al.}(2015){Bacon}, {Brinchmann}, {Richard}, {Contini},
  {Drake}, {Franx}, {Tacchella}, {Vernet}, {Wisotzki}, {Blaizot}, {Bouch{\'e}},
  {Bouwens}, {Cantalupo}, {Carollo}, {Carton}, {Caruana}, {Cl{\'e}ment},
  {Dreizler}, {Epinat}, {Guiderdoni}, {Herenz}, {Husser}, {Kamann}, {Kerutt},
  {Kollatschny}, {Krajnovic}, {Lilly}, {Martinsson}, {Michel-Dansac},
  {Patricio}, {Schaye}, {Shirazi}, {Soto}, {Soucail}, {Steinmetz}, {Urrutia},
  {Weilbacher}, \& {de Zeeuw}}]{Bacon2015}
{Bacon}, R., {Brinchmann}, J., {Richard}, J., {et~al.} 2015, AAP, 575, A75

\bibitem[{{Bouwens} {et~al.}(2009){Bouwens}, {Illingworth}, {Franx}, {Chary},
  {Meurer}, {Conselice}, {Ford}, {Giavalisco}, \& {van Dokkum}}]{Bouwens2009}
{Bouwens}, R.~J., {Illingworth}, G.~D., {Franx}, M., {et~al.} 2009, ApJ, 705,
  936

\bibitem[{{Cai} {et~al.}(2015){Cai}, {Fan}, {Jiang}, {Dav{\'e}}, {Oh}, {Yang},
  \& {Zabludoff}}]{Cai2015}
{Cai}, Z., {Fan}, X., {Jiang}, L., {et~al.} 2015, \apjl, 799, L19

\bibitem[{{Calzetti} {et~al.}(2000){Calzetti}, {Armus}, {Bohlin}, {Kinney},
  {Koornneef}, \& {Storchi-Bergmann}}]{Calzetti00}
{Calzetti}, D., {Armus}, L., {Bohlin}, R.~C., {et~al.} 2000, ApJ, 533, 682

\bibitem[{{Cardamone} {et~al.}(2009)}]{Cardamone2009}
{Cardamone}, C. {et~al.} 2009, MNRAS, 399, 1191

\bibitem[{{Caruana} {et~al.}(2018){Caruana}, {Wisotzki}, {Herenz},
  {et~al.}}]{Caruana2018}
{Caruana}, J., {Wisotzki}, L., {Herenz}, E.~C., {et~al.} 2018, \mnras, 473, 30

\bibitem[{{Cassata} {et~al.}(2015){Cassata}, {Tasca}, {Le F{\`e}vre},
  {et~al.}}]{Cassata2015}
{Cassata}, P., {Tasca}, L.~A.~M., {Le F{\`e}vre}, O., {et~al.} 2015, \aap, 573,
  A24

\bibitem[{{Cassata} {et~al.}(2011)}]{Cassata2011}
{Cassata}, P. {et~al.} 2011, A\&A, 525, A143

\bibitem[{{Charlot} \& {Fall}(1993)}]{Charlot1993}
{Charlot}, S. \& {Fall}, S.~M. 1993, \apj, 415, 580

\bibitem[{{Chisholm} {et~al.}(2018){Chisholm}, {Gazagnes}, {Schaerer},
  {Verhamme}, {Rigby}, {Bayliss}, {Sharon}, {Gladders}, \&
  {Dahle}}]{Chisholm2018}
{Chisholm}, J., {Gazagnes}, S., {Schaerer}, D., {et~al.} 2018, \aap, 616, A30

\bibitem[{{Ciardullo} {et~al.}(2014){Ciardullo}, {Zeimann}, {Gronwall},
  {Gebhardt}, {Schneider}, {Hagen}, {Malz}, {Blanc}, {Hill}, {Drory}, \&
  {Gawiser}}]{Ciardullo2014}
{Ciardullo}, R., {Zeimann}, G.~R., {Gronwall}, C., {et~al.} 2014, \apj, 796, 64

\bibitem[{{Dijkstra}(2017)}]{Dijkstra2017}
{Dijkstra}, M. 2017, ArXiv e-prints [\eprint[arXiv]{1704.03416}]

\bibitem[{{Dijkstra} \& {Westra}(2010)}]{Dijkstra2010}
{Dijkstra}, M. \& {Westra}, E. 2010, MNRAS, 401, 2343

\bibitem[{{Ding} {et~al.}(2017){Ding}, {Cai}, {Fan}, {Stark}, {Bian}, {Jiang},
  {McGreer}, {Robertson}, \& {Siana}}]{Ding2017}
{Ding}, J., {Cai}, Z., {Fan}, X., {et~al.} 2017, \apjl, 838, L22

\bibitem[{{Dom{\'{\i}}nguez S{\'a}nchez} {et~al.}(2012)}]{Dom2012}
{Dom{\'{\i}}nguez S{\'a}nchez}, H. {et~al.} 2012, \mnras, 426, 330

\bibitem[{{Drake} {et~al.}(2017{\natexlab{a}})}]{Drake2017a}
{Drake}, A.~B. {et~al.} 2017{\natexlab{a}}, \mnras, 471, 267

\bibitem[{{Drake} {et~al.}(2017{\natexlab{b}})}]{Drake2017b}
{Drake}, A.~B. {et~al.} 2017{\natexlab{b}}, \aap, 608, A6

\bibitem[{{Dressler} {et~al.}(2015){Dressler}, {Henry}, {Martin}, {Sawicki},
  {McCarthy}, \& {Villaneuva}}]{Dressler2015}
{Dressler}, A., {Henry}, A., {Martin}, C.~L., {et~al.} 2015, \apj, 806, 19

\bibitem[{{Fletcher} {et~al.}(2018){Fletcher}, {Robertson}, {Nakajima},
  {Ellis}, {Stark}, \& {Inoue}}]{Fletcher2018}
{Fletcher}, T.~J., {Robertson}, B.~E., {Nakajima}, K., {et~al.} 2018, ArXiv
  e-prints [\eprint[arXiv]{1806.01741}]

\bibitem[{{Fynbo} {et~al.}(2003){Fynbo}, {Ledoux}, {M{\o}ller}, {Thomsen}, \&
  {Burud}}]{Fynbo2003}
{Fynbo}, J.~P.~U., {Ledoux}, C., {M{\o}ller}, P., {Thomsen}, B., \& {Burud}, I.
  2003, \aap, 407, 147

\bibitem[{{Garn} \& {Best}(2010)}]{GarnBest2010}
{Garn}, T. \& {Best}, P.~N. 2010, MNRAS, 409, 421

\bibitem[{{Gawiser} {et~al.}(2007){Gawiser}, {Francke}, {Lai},
  {et~al.}}]{Gawiser2007}
{Gawiser}, E., {Francke}, H., {Lai}, K., {et~al.} 2007, \apj, 671, 278

\bibitem[{{Gronke} {et~al.}(2015){Gronke}, {Dijkstra}, {Trenti}, \&
  {Wyithe}}]{GronkeDi2015}
{Gronke}, M., {Dijkstra}, M., {Trenti}, M., \& {Wyithe}, S. 2015, \mnras, 449,
  1284

\bibitem[{{Hagen} {et~al.}(2016){Hagen}, {Zeimann}, {et~al.}}]{Hagen2016}
{Hagen}, A., {Zeimann}, G.~R., {et~al.} 2016, \apj, 817, 79

\bibitem[{{Harikane} {et~al.}(2018){Harikane}, {Ouchi}, {Shibuya},
  {et~al.}}]{Harikane2017}
{Harikane}, Y., {Ouchi}, M., {Shibuya}, T., {et~al.} 2018, \apj, 859, 84

\bibitem[{{Hashimoto} {et~al.}(2017)}]{Hashimoto2017}
{Hashimoto}, T. {et~al.} 2017, \aap, 608, A10

\bibitem[{{Hayes} {et~al.}(2013){Hayes}, {{\"O}stlin}, {Schaerer}, {Verhamme},
  {Mas-Hesse}, {Adamo}, {Atek}, {Cannon}, {Duval}, {Guaita}, {Herenz}, {Kunth},
  {Laursen}, {Melinder}, {Orlitov{\'a}}, {Ot{\'{\i}}-Floranes}, \&
  {Sandberg}}]{Hayes2013}
{Hayes}, M., {{\"O}stlin}, G., {Schaerer}, D., {et~al.} 2013, ApJL, 765, L27

\bibitem[{{Hayes} {et~al.}(2011){Hayes}, {Schaerer}, {{\"O}stlin}, {Mas-Hesse},
  {Atek}, \& {Kunth}}]{Hayes2011}
{Hayes}, M., {Schaerer}, D., {{\"O}stlin}, G., {et~al.} 2011, ApJ, 730, 8

\bibitem[{{Hayes} {et~al.}(2010)}]{Hayes2010}
{Hayes}, M. {et~al.} 2010, Nature, 464, 562

\bibitem[{{Hayes} {et~al.}(2014)}]{Hayes2014}
{Hayes}, M. {et~al.} 2014, ApJ, 782, 6

\bibitem[{{Henry} {et~al.}(2015){Henry}, {Scarlata}, {Martin}, \&
  {Erb}}]{Henry2015}
{Henry}, A., {Scarlata}, C., {Martin}, C.~L., \& {Erb}, D. 2015, \apj, 809, 19

\bibitem[{{Herenz} {et~al.}(2017){Herenz}, {Urrutia}, {Wisotzki},
  {et~al.}}]{Herenz2017}
{Herenz}, E.~C., {Urrutia}, T., {Wisotzki}, L., {et~al.} 2017, \aap, 606, A12

\bibitem[{{Hu} {et~al.}(2010){Hu}, {Cowie}, {Barger}, {Capak}, {Kakazu}, \&
  {Trouille}}]{Hu2010}
{Hu}, E.~M., {Cowie}, L.~L., {Barger}, A.~J., {et~al.} 2010, ApJ, 725, 394

\bibitem[{{Hu} {et~al.}(2016){Hu}, {Cowie}, {Songaila}, {Barger},
  {Rosenwasser}, \& {Wold}}]{Hu2016}
{Hu}, E.~M., {Cowie}, L.~L., {Songaila}, A., {et~al.} 2016, ApJL, 825, L7

\bibitem[{Hunter(2007)}]{Hunter:2007}
Hunter, J.~D. 2007, Computing In Science \& Engineering, 9, 90

\bibitem[{{Izotov} {et~al.}(2016{\natexlab{a}}){Izotov}, {Orlitov{\'a}},
  {Schaerer}, {Thuan}, {Verhamme}, {Guseva}, \& {Worseck}}]{Izotov2016a}
{Izotov}, Y.~I., {Orlitov{\'a}}, I., {Schaerer}, D., {et~al.}
  2016{\natexlab{a}}, Nature, 529, 178

\bibitem[{{Izotov} {et~al.}(2016{\natexlab{b}}){Izotov}, {Schaerer}, {Thuan},
  {Worseck}, {Guseva}, {Orlitov{\'a}}, \& {Verhamme}}]{Izotov2016b}
{Izotov}, Y.~I., {Schaerer}, D., {Thuan}, T.~X., {et~al.} 2016{\natexlab{b}},
  MNRAS, 461, 3683

\bibitem[{{Izotov} {et~al.}(2018){Izotov}, {Schaerer}, {Worseck}, {Guseva},
  {Thuan}, {Verhamme}, {Orlitov{\'a}}, \& {Fricke}}]{Izotov2017_Lya}
{Izotov}, Y.~I., {Schaerer}, D., {Worseck}, G., {et~al.} 2018, \mnras, 474,
  4514

\bibitem[{Jones {et~al.}(2001)Jones, Oliphant, Peterson, {et~al.}}]{jones}
Jones, E., Oliphant, T., Peterson, P., {et~al.} 2001, {SciPy}: Open source
  scientific tools for {Python}

\bibitem[{{Kennicutt}(1998)}]{Kennicutt1998}
{Kennicutt}, Jr., R.~C. 1998, ARAA, 36, 189

\bibitem[{{Kusakabe} {et~al.}(2018){Kusakabe}, {Shimasaku}, {Ouchi},
  {Nakajima}, {Goto}, {Hashimoto}, {Konno}, {Harikane}, {Silverman}, \&
  {Capak}}]{Kusakabe2018}
{Kusakabe}, H., {Shimasaku}, K., {Ouchi}, M., {et~al.} 2018, \pasj, 70, 4

\bibitem[{{Laursen} {et~al.}(2011){Laursen}, {Sommer-Larsen}, \&
  {Razoumov}}]{Laursen2011}
{Laursen}, P., {Sommer-Larsen}, J., \& {Razoumov}, A.~O. 2011, \apj, 728, 52

\bibitem[{{Mainali} {et~al.}(2017){Mainali}, {Kollmeier}, {Stark}, {Simcoe},
  {Walth}, {Newman}, \& {Miller}}]{Mainali2017}
{Mainali}, R., {Kollmeier}, J.~A., {Stark}, D.~P., {et~al.} 2017, \apjl, 836,
  L14

\bibitem[{{Malhotra} \& {Rhoads}(2004)}]{Malhotra2004}
{Malhotra}, S. \& {Rhoads}, J.~E. 2004, ApJL, 617, L5

\bibitem[{{Martin} \& {Sawicki}(2004)}]{Martin2004}
{Martin}, C.~L. \& {Sawicki}, M. 2004, ApJ, 603, 414

\bibitem[{{Maseda} {et~al.}(2018)}]{Maseda2018}
{Maseda}, M.~V. {et~al.} 2018, \apjl, 865, L1

\bibitem[{{Mason} {et~al.}(2018{\natexlab{a}}){Mason}, {Treu}, {de Barros},
  {Dijkstra}, {Fontana}, {Mesinger}, {Pentericci}, {Trenti}, \&
  {Vanzella}}]{Mason2018a}
{Mason}, C.~A., {Treu}, T., {de Barros}, S., {et~al.} 2018{\natexlab{a}},
  \apjl, 857, L11

\bibitem[{{Mason} {et~al.}(2018{\natexlab{b}}){Mason}, {Treu}, {Dijkstra},
  {Mesinger}, {Trenti}, {Pentericci}, {de Barros}, \& {Vanzella}}]{Mason2018b}
{Mason}, C.~A., {Treu}, T., {Dijkstra}, M., {et~al.} 2018{\natexlab{b}}, \apj,
  856, 2

\bibitem[{{Matthee} {et~al.}(2017{\natexlab{a}}){Matthee}, {Sobral}, {Best},
  {Khostovan}, {Oteo}, {Bouwens}, \& {R{\"o}ttgering}}]{MattheeGALEX2017}
{Matthee}, J., {Sobral}, D., {Best}, P., {et~al.} 2017{\natexlab{a}}, \mnras,
  465, 3637

\bibitem[{{Matthee} {et~al.}(2017{\natexlab{b}}){Matthee}, {Sobral}, {Boone},
  {et~al.}}]{Matthee_CR7_17}
{Matthee}, J., {Sobral}, D., {Boone}, F., {et~al.} 2017{\natexlab{b}}, \apj,
  851, 145

\bibitem[{{Matthee} {et~al.}(2017{\natexlab{c}}){Matthee}, {Sobral}, {Darvish},
  {Santos}, {Mobasher}, {Paulino-Afonso}, {R{\"o}ttgering}, \&
  {Alegre}}]{Matthee2017followup}
{Matthee}, J., {Sobral}, D., {Darvish}, B., {et~al.} 2017{\natexlab{c}},
  \mnras, 472, 772

\bibitem[{{Matthee} {et~al.}(2018){Matthee}, {Sobral}, {Gronke},
  {Paulino-Afonso}, {Stefanon}, \& {R{\"o}ttgering}}]{MattheeCOLA1}
{Matthee}, J., {Sobral}, D., {Gronke}, M., {et~al.} 2018, \aap, 619, A136

\bibitem[{{Matthee} {et~al.}(2016){Matthee}, {Sobral}, {Oteo}, {Best}, {Smail},
  {R{\"o}ttgering}, \& {Paulino-Afonso}}]{Matthee2016}
{Matthee}, J., {Sobral}, D., {Oteo}, I., {et~al.} 2016, \mnras, 458, 449

\bibitem[{{Matthee} {et~al.}(2015){Matthee}, {Sobral}, {Santos},
  {R{\"o}ttgering}, {Darvish}, \& {Mobasher}}]{Matthee2015}
{Matthee}, J., {Sobral}, D., {Santos}, S., {et~al.} 2015, MNRAS, 451, 400

\bibitem[{{Meurer} {et~al.}(1999){Meurer}, {Heckman}, \&
  {Calzetti}}]{Meurer1999}
{Meurer}, G.~R., {Heckman}, T.~M., \& {Calzetti}, D. 1999, ApJ, 521, 64

\bibitem[{{Miley} \& {De Breuck}(2008)}]{Miley2008}
{Miley}, G. \& {De Breuck}, C. 2008, AAPR, 15, 67

\bibitem[{{Momose} {et~al.}(2014){Momose}, {Ouchi}, {Nakajima}, {Ono},
  {Shibuya}, {Shimasaku}, {Yuma}, {Mori}, \& {Umemura}}]{Momose2014}
{Momose}, R., {Ouchi}, M., {Nakajima}, K., {et~al.} 2014, \mnras, 442, 110

\bibitem[{{Nakajima} {et~al.}(2018){Nakajima}, {Fletcher}, {Ellis},
  {Robertson}, \& {Iwata}}]{Nakajima2018}
{Nakajima}, K., {Fletcher}, T., {Ellis}, R.~S., {Robertson}, B.~E., \& {Iwata},
  I. 2018, \mnras, 477, 2098

\bibitem[{{Nakajima} \& {Ouchi}(2014)}]{NakajimaOuchi2013}
{Nakajima}, K. \& {Ouchi}, M. 2014, MNRAS, 442, 900

\bibitem[{{Narayanan} {et~al.}(2018){Narayanan}, {Dav{\'e}}, {Johnson},
  {Thompson}, {Conroy}, \& {Geach}}]{Narayanan2018}
{Narayanan}, D., {Dav{\'e}}, R., {Johnson}, B.~D., {et~al.} 2018, \mnras, 474,
  1718

\bibitem[{{Neufeld}(1991)}]{Neufeld1991}
{Neufeld}, D.~A. 1991, \apjl, 370, L85

\bibitem[{{Oesch} {et~al.}(2015){Oesch}, {van Dokkum}, {Illingworth},
  {Bouwens}, {Momcheva}, {Holden}, {Roberts-Borsani}, {Smit}, {Franx},
  {Labb{\'e}}, {Gonz{\'a}lez}, \& {Magee}}]{Oesch2015}
{Oesch}, P.~A., {van Dokkum}, P.~G., {Illingworth}, G.~D., {et~al.} 2015,
  \apjl, 804, L30

\bibitem[{{Oke} \& {Gunn}(1983)}]{Oke1983}
{Oke}, J.~B. \& {Gunn}, J.~E. 1983, \apj, 266, 713

\bibitem[{{Ono} {et~al.}(2012)}]{Ono2012}
{Ono}, Y. {et~al.} 2012, ApJ, 744, 83

\bibitem[{{Oteo} {et~al.}(2015){Oteo}, {Sobral}, {Ivison}, {Smail}, {Best},
  {Cepa}, \& {P{\'e}rez-Garc{\'{\i}}a}}]{Oteo2015}
{Oteo}, I., {Sobral}, D., {Ivison}, R.~J., {et~al.} 2015, \mnras, 452, 2018

\bibitem[{{Ouchi} {et~al.}(2008)}]{Ouchi2008}
{Ouchi}, M. {et~al.} 2008, ApJs, 176, 301

\bibitem[{{Ouchi} {et~al.}(2009)}]{Ouchi2009}
{Ouchi}, M. {et~al.} 2009, ApJ, 696, 1164

\bibitem[{{Oyarz{\'u}n} {et~al.}(2017){Oyarz{\'u}n}, {Blanc}, {Gonz{\'a}lez},
  {Mateo}, \& {Bailey}}]{Oyarzun2017}
{Oyarz{\'u}n}, G.~A., {Blanc}, G.~A., {Gonz{\'a}lez}, V., {Mateo}, M., \&
  {Bailey}, III, J.~I. 2017, \apj, 843, 133

\bibitem[{{Partridge} \& {Peebles}(1967)}]{Partridge1967}
{Partridge}, R.~B. \& {Peebles}, P.~J.~E. 1967, ApJ, 147, 868

\bibitem[{{Paulino-Afonso} {et~al.}(2018){Paulino-Afonso}, {Sobral}, {Ribeiro},
  {Matthee}, {Santos}, {Calhau}, {Forshaw}, {Johnson}, {Merrick}, {Perez}, \&
  {Sheldon}}]{Paulino-Afonso2017}
{Paulino-Afonso}, A., {Sobral}, D., {Ribeiro}, B., {et~al.} 2018, \mnras, 476,
  5479

\bibitem[{{Popping} {et~al.}(2017){Popping}, {Puglisi}, \&
  {Norman}}]{Popping2017}
{Popping}, G., {Puglisi}, A., \& {Norman}, C.~A. 2017, \mnras, 472, 2315

\bibitem[{{Pritchet}(1994)}]{Pritchet1994}
{Pritchet}, C.~J. 1994, PASP, 106, 1052

\bibitem[{{Rauch} {et~al.}(2008)}]{Rauch2008}
{Rauch}, M. {et~al.} 2008, \apj, 681, 856

\bibitem[{{Reddy} {et~al.}(2015){Reddy}, {Kriek}, {Shapley}, {Freeman},
  {Siana}, {Coil}, {Mobasher}, {Price}, {Sanders}, \& {Shivaei}}]{Reddy2015}
{Reddy}, N.~A., {Kriek}, M., {Shapley}, A.~E., {et~al.} 2015, ApJ, 806, 259

\bibitem[{{Reddy} {et~al.}(2016){Reddy}, {Steidel}, {Pettini},
  {Bogosavljevi{\'c}}, \& {Shapley}}]{Reddy2016}
{Reddy}, N.~A., {Steidel}, C.~C., {Pettini}, M., {Bogosavljevi{\'c}}, M., \&
  {Shapley}, A.~E. 2016, \apj, 828, 108

\bibitem[{{Rhoads} {et~al.}(2000){Rhoads}, {Malhotra}, {Dey}, {Stern},
  {Spinrad}, \& {Jannuzi}}]{Rhoads2000}
{Rhoads}, J.~E., {Malhotra}, S., {Dey}, A., {et~al.} 2000, ApJL, 545, L85

\bibitem[{{Salpeter}(1955)}]{Salpeter1955}
{Salpeter}, E.~E. 1955, \apj, 121, 161

\bibitem[{{Schaerer}(2003)}]{Schaerer2003}
{Schaerer}, D. 2003, AAP, 397, 527

\bibitem[{{Schmidt} {et~al.}(2017)}]{Schmidt2017}
{Schmidt}, K.~B. {et~al.} 2017, ApJ, 839, 17

\bibitem[{{Shibuya} {et~al.}(2018)}]{Shibuya2017_spec}
{Shibuya}, T. {et~al.} 2018, \pasj, 70, S15

\bibitem[{{Shivaei} {et~al.}(2018){Shivaei}, {Reddy}, {et~al.}}]{Shivaei2017}
{Shivaei}, I., {Reddy}, N.~A., {et~al.} 2018, \apj, 855, 42

\bibitem[{{Smidt} {et~al.}(2016){Smidt}, {Wiggins}, \& {Johnson}}]{Smidt2016}
{Smidt}, J., {Wiggins}, B.~K., \& {Johnson}, J.~L. 2016, \apjl, 829, L6

\bibitem[{{Smit} {et~al.}(2012){Smit}, {Bouwens}, {Franx}, {Illingworth},
  {Labb{\'e}}, {Oesch}, \& {van Dokkum}}]{Smit2012}
{Smit}, R., {Bouwens}, R.~J., {Franx}, M., {et~al.} 2012, \apj, 756, 14

\bibitem[{{Sobral} {et~al.}(2012){Sobral}, {Best}, {Matsuda}, {Smail}, {Geach},
  \& {Cirasuolo}}]{Sobral2012}
{Sobral}, D., {Best}, P.~N., {Matsuda}, Y., {et~al.} 2012, MNRAS, 420, 1926

\bibitem[{{Sobral} {et~al.}(2014){Sobral}, {Best}, {Smail}, {Mobasher},
  {Stott}, \& {Nisbet}}]{Sobral2014}
{Sobral}, D., {Best}, P.~N., {Smail}, I., {et~al.} 2014, \mnras, 437, 3516

\bibitem[{{Sobral} {et~al.}(2017){Sobral}, {Matthee}, {Best}, {Stroe},
  {R{\"o}ttgering}, {Oteo}, {Smail}, {Morabito}, \&
  {Paulino-Afonso}}]{Sobral2017}
{Sobral}, D., {Matthee}, J., {Best}, P., {et~al.} 2017, \mnras, 466, 1242

\bibitem[{{Sobral} {et~al.}(2015){Sobral}, {Matthee}, {Darvish}, {Schaerer},
  {Mobasher}, {R{\"o}ttgering}, {Santos}, \& {Hemmati}}]{Sobral2015}
{Sobral}, D., {Matthee}, J., {Darvish}, B., {et~al.} 2015, ApJ, 808, 139

\bibitem[{{Sobral} {et~al.}(2018{\natexlab{a}}){Sobral}, {Santos}, {Matthee},
  {Paulino-Afonso}, {Ribeiro}, {Calhau}, \& {Khostovan}}]{Sobral2017_SC4K}
{Sobral}, D., {Santos}, S., {Matthee}, J., {et~al.} 2018{\natexlab{a}}, \mnras,
  476, 4725

\bibitem[{{Sobral} {et~al.}(2018{\natexlab{b}})}]{Sobral2018b}
{Sobral}, D. {et~al.} 2018{\natexlab{b}}, \mnras, 477, 2817

\bibitem[{{Song} {et~al.}(2014){Song}, {Finkelstein}, {et~al.}}]{Song2014}
{Song}, M., {Finkelstein}, S.~L., {et~al.} 2014, \apj, 791, 3

\bibitem[{{Stark} {et~al.}(2017){Stark}, {Ellis}, {Charlot},
  {et~al.}}]{Stark2017}
{Stark}, D.~P., {Ellis}, R.~S., {Charlot}, S., {et~al.} 2017, \mnras, 464, 469

\bibitem[{{Stark} {et~al.}(2010){Stark}, {Ellis}, {Chiu}, {Ouchi}, \&
  {Bunker}}]{Stark2010}
{Stark}, D.~P., {Ellis}, R.~S., {Chiu}, K., {Ouchi}, M., \& {Bunker}, A. 2010,
  MNRAS, 408, 1628

\bibitem[{{Stark} {et~al.}(2015{\natexlab{a}}){Stark}, {Richard}, {Charlot},
  {et~al.}}]{Stark2015a}
{Stark}, D.~P., {Richard}, J., {Charlot}, S., {et~al.} 2015{\natexlab{a}},
  \mnras, 450, 1846

\bibitem[{{Stark} {et~al.}(2015{\natexlab{b}}){Stark}, {Walth}, {Charlot},
  {et~al.}}]{Stark2015b}
{Stark}, D.~P., {Walth}, G., {Charlot}, S., {et~al.} 2015{\natexlab{b}},
  \mnras, 454, 1393

\bibitem[{{Stark} {et~al.}(2015{\natexlab{c}})}]{Stark2015}
{Stark}, D.~P. {et~al.} 2015{\natexlab{c}}, MNRAS, 450, 1846

\bibitem[{{Steidel} {et~al.}(2018){Steidel}, {Bogosavlevic}, {Shapley},
  {Reddy}, {Rudie}, {Pettini}, {Trainor}, \& {Strom}}]{Steidel2018}
{Steidel}, C.~C., {Bogosavlevic}, M., {Shapley}, A.~E., {et~al.} 2018, \apj,
  869, 123

\bibitem[{{Steidel} {et~al.}(2011){Steidel}, {Bogosavljevi{\'c}}, {Shapley},
  {Kollmeier}, {Reddy}, {Erb}, \& {Pettini}}]{Steidel2011}
{Steidel}, C.~C., {Bogosavljevi{\'c}}, M., {Shapley}, A.~E., {et~al.} 2011,
  \apj, 736, 160

\bibitem[{{Steidel} {et~al.}(2016){Steidel}, {Strom}, {Pettini}, {Rudie},
  {Reddy}, \& {Trainor}}]{Steidel2016}
{Steidel}, C.~C., {Strom}, A.~L., {Pettini}, M., {et~al.} 2016, \apj, 826, 159

\bibitem[{{Suzuki} {et~al.}(2017)}]{Suzuki2017}
{Suzuki}, T.~L. {et~al.} 2017, ApJ, 849, 39

\bibitem[{{Taylor}(2013)}]{Topcat}
{Taylor}, M. 2013, Starlink User Note, 253

\bibitem[{{Tilvi} {et~al.}(2016){Tilvi}, {Pirzkal}, {Malhotra},
  {et~al.}}]{Tilvi2016}
{Tilvi}, V., {Pirzkal}, N., {Malhotra}, S., {et~al.} 2016, \apjl, 827, L14

\bibitem[{{Trainor} {et~al.}(2015){Trainor}, {Steidel}, {Strom}, \&
  {Rudie}}]{Trainor2015}
{Trainor}, R.~F., {Steidel}, C.~C., {Strom}, A.~L., \& {Rudie}, G.~C. 2015,
  \apj, 809, 89

\bibitem[{{Trainor} {et~al.}(2016){Trainor}, {Strom}, {Steidel}, \&
  {Rudie}}]{Trainor2016}
{Trainor}, R.~F., {Strom}, A.~L., {Steidel}, C.~C., \& {Rudie}, G.~C. 2016,
  \apj, 832, 171

\bibitem[{{van Breukelen} {et~al.}(2005){van Breukelen}, {Jarvis}, \&
  {Venemans}}]{vanBreukelen2005}
{van Breukelen}, C., {Jarvis}, M.~J., \& {Venemans}, B.~P. 2005, \mnras, 359,
  895

\bibitem[{Van Der~Walt {et~al.}(2011)Van Der~Walt, Colbert, \&
  Varoquaux}]{van2011numpy}
Van Der~Walt, S., Colbert, S.~C., \& Varoquaux, G. 2011, Computing in Science
  \& Engineering, 13, 22

\bibitem[{{Vanzella} {et~al.}(2011){Vanzella}, {Pentericci}, {Fontana},
  {et~al.}}]{Vanzella2011}
{Vanzella}, E., {Pentericci}, L., {Fontana}, A., {et~al.} 2011, \apjl, 730, L35

\bibitem[{{Verhamme} {et~al.}(2017){Verhamme}, {Orlitov{\'a}}, {Schaerer},
  {Izotov}, {Worseck}, {Thuan}, \& {Guseva}}]{Verhamme2017}
{Verhamme}, A., {Orlitov{\'a}}, I., {Schaerer}, D., {et~al.} 2017, \aap, 597,
  A13

\bibitem[{{Verhamme} {et~al.}(2008){Verhamme}, {Schaerer}, {Atek}, \&
  {Tapken}}]{Verhamme2008}
{Verhamme}, A., {Schaerer}, D., {Atek}, H., \& {Tapken}, C. 2008, \aap, 491, 89

\bibitem[{{Wilkins} {et~al.}(2013){Wilkins}, {Bunker}, {Coulton}, {Croft}, {di
  Matteo}, {Khandai}, \& {Feng}}]{Wilkins2013}
{Wilkins}, S.~M., {Bunker}, A., {Coulton}, W., {et~al.} 2013, \mnras, 430, 2885

\bibitem[{{Wisotzki} {et~al.}(2016){Wisotzki}, {Bacon},
  {et~al.}}]{Wisotzki2016}
{Wisotzki}, L., {Bacon}, R., {et~al.} 2016, \aap, 587, A98

\bibitem[{{Yang} {et~al.}(2016){Yang}, {Malhotra}, {Gronke}, {Rhoads},
  {Dijkstra}, {Jaskot}, {Zheng}, \& {Wang}}]{Yang2016}
{Yang}, H., {Malhotra}, S., {Gronke}, M., {et~al.} 2016, ApJ, 820, 130

\bibitem[{{Yang} {et~al.}(2017){Yang}, {Malhotra}, {Gronke}, {Rhoads},
  {Leitherer}, {Wofford}, {Jiang}, {Dijkstra}, {Tilvi}, \& {Wang}}]{Yang2017}
{Yang}, H., {Malhotra}, S., {Gronke}, M., {et~al.} 2017, \apj, 844, 171

\bibitem[{{Zabl} {et~al.}(2015){Zabl}, {N{\o}rgaard-Nielsen}, {Fynbo},
  {Laursen}, {Ouchi}, \& {Kj{\ae}rgaard}}]{Zabl2015}
{Zabl}, J., {N{\o}rgaard-Nielsen}, H.~U., {Fynbo}, J.~P.~U., {et~al.} 2015,
  MNRAS, 451, 2050

\end{thebibliography}

\begin{appendix}

\section{Toy-model grid for $\rm f_{esc,Ly\alpha}$ dependencies}\label{Toy_model}

We construct a simple analytical toy-model grid to produce observable H$\alpha$, UV and Ly$\alpha$ luminosities and EW$_0$ from a range of input physical conditions (see Table \ref{Params_toy_model}). We independently sample in steps of 0.01 or 0.01\,dex combinations of SFR, $\rm f_{esc,LyC}$, case B Ly$\alpha$/H$\alpha$ intrinsic ratio, $\rm \log_{10}(\xi_{ion}/Hz\,erg^{-1})$, $E(B-V)$ with a \cite{Calzetti00} dust attenuation law and a parameter to control $\rm f_{esc,Ly\alpha}$ (from e.g. scattering leading to higher dust absorption or scattering Ly$\alpha$ photons away from or into the observers' line of sight) which acts as a further factor affecting $\rm f_{esc,Ly\alpha}$; see Table \ref{Params_toy_model} for the range in parameters explored independently. We follow \cite{Kennicutt1998} and all definitions and assumptions mentioned in this paper and we sample the parameter space with a flat prior. We use the \cite{Meurer1999} relation: $\rm A_{UV}\approx4.43+1.99\beta$ (based on an intrinsic slope $\beta_{int}=-2.23$), but we also vary the intrinsic $\beta_{int}$ from $-2.6$ to $-1.8$ \citep[see e.g.][]{Wilkins2013}. We publicly release our simple {\sc python} script which can be used for similar studies and/or to study different ranges in the parameter space, or conduct studies in which properties are intrinsically related/linked as one expects for realistic galaxies. Section \ref{equations_grid} presents the full equations implemented in the publicly available {\sc python} script.

%%%%%%%%%%%%%%%%%%%%%%%%%%%%%%%%%%%%%%%%%%%
%
% Table A1 - Parameters used for the Toy Model
%
%%%%%%%%%%%%%%%%%%%%%%%%%%%%%%%%%%%%%%%%%%%
\begin{table}
 \centering
\caption{The parameters varied in our simple toy model to produce a grid of 1,000,000 sources to interpret the observational results (see Appendix \ref{Toy_model}).}
 \label{Params_toy_model}
\begin{tabular}{ccccc}
\hline
Property  & Minimum  & Maximum & $\Delta$ param. & \\ 
\hline
SFR (M$_{\odot}$\,yr$^{-1}$)  &  0.1 & 100  & 0.01\,dex \\ 
$\rm \log_{10}(\xi_{ion}/Hz\,erg^{-1})$  &  24.7 & 26.5  & 0.01\,dex \\ 
$\rm f_{esc,LyC}$  &  0.0 & 0.15  & 0.01 \\ 
Ly$\alpha$/H$\alpha$  & 8.0 & 9.0  & 0.01 \\ 
$E(B-V)$ &  0.0  & 0.5  & 0.01 \\ 
$\rm \beta_{int}$ &  $-2.6$ & $-1.8$  &  0.01 \\ 
Extra $\rm f_{esc,Ly\alpha}$ &  0.0 & 1.3  & 0.01 \\ 
\hline
\end{tabular}
\end{table}

%%%%%%%%%%%%%%%%%%%%%%%%%%%%%%%%%%%%%%%%%%%%%%%%
%
% FIGURE A1 - Further Physical Interpretation
%
%%%%%%%%%%%%%%%%%%%%%%%%%%%%%%%%%%%%%%%%%%%%%%%%
\begin{figure*}
\begin{tabular}{cc}
\includegraphics[width=8.88cm]{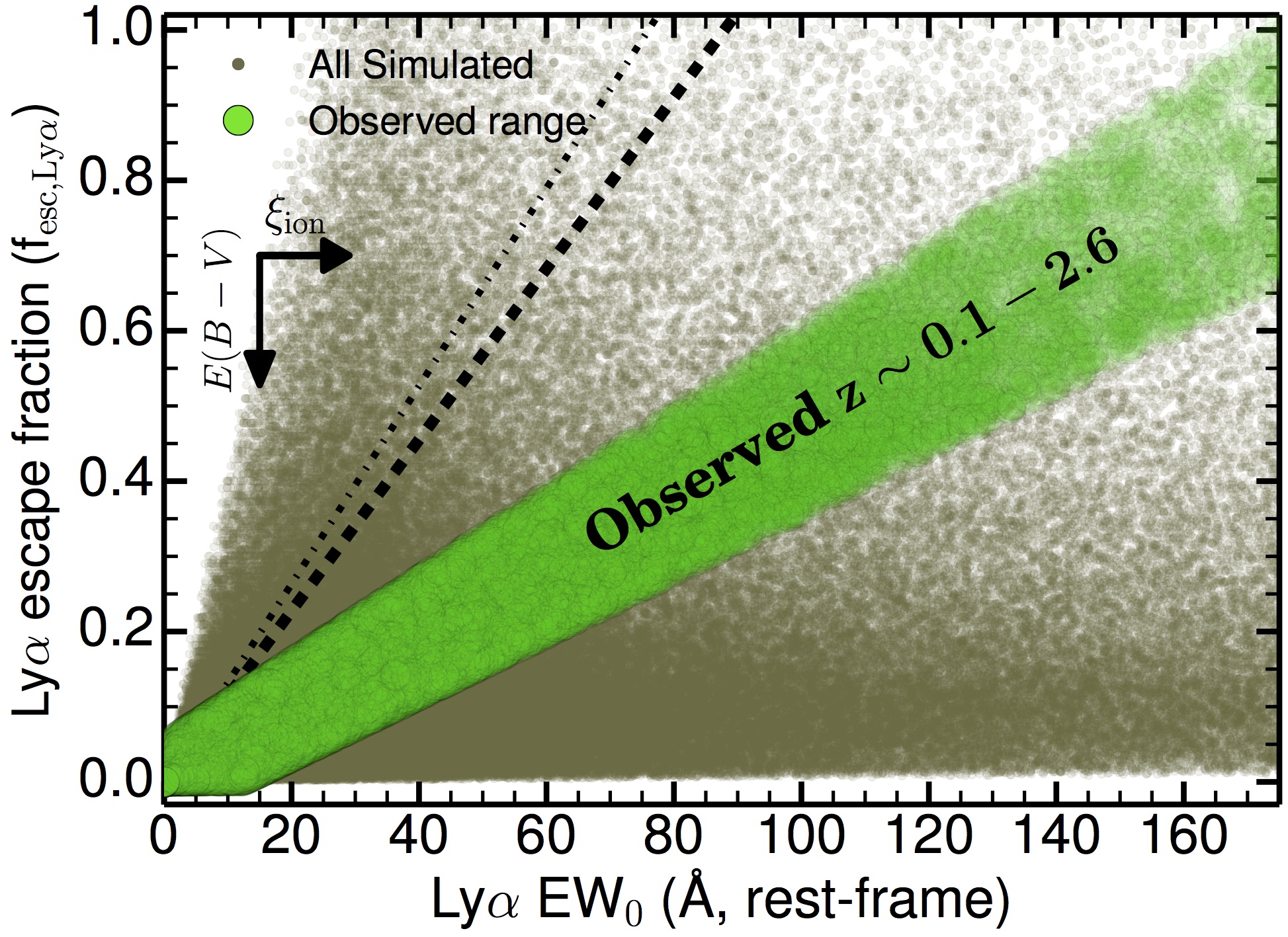}&  
\includegraphics[width=8.76cm]{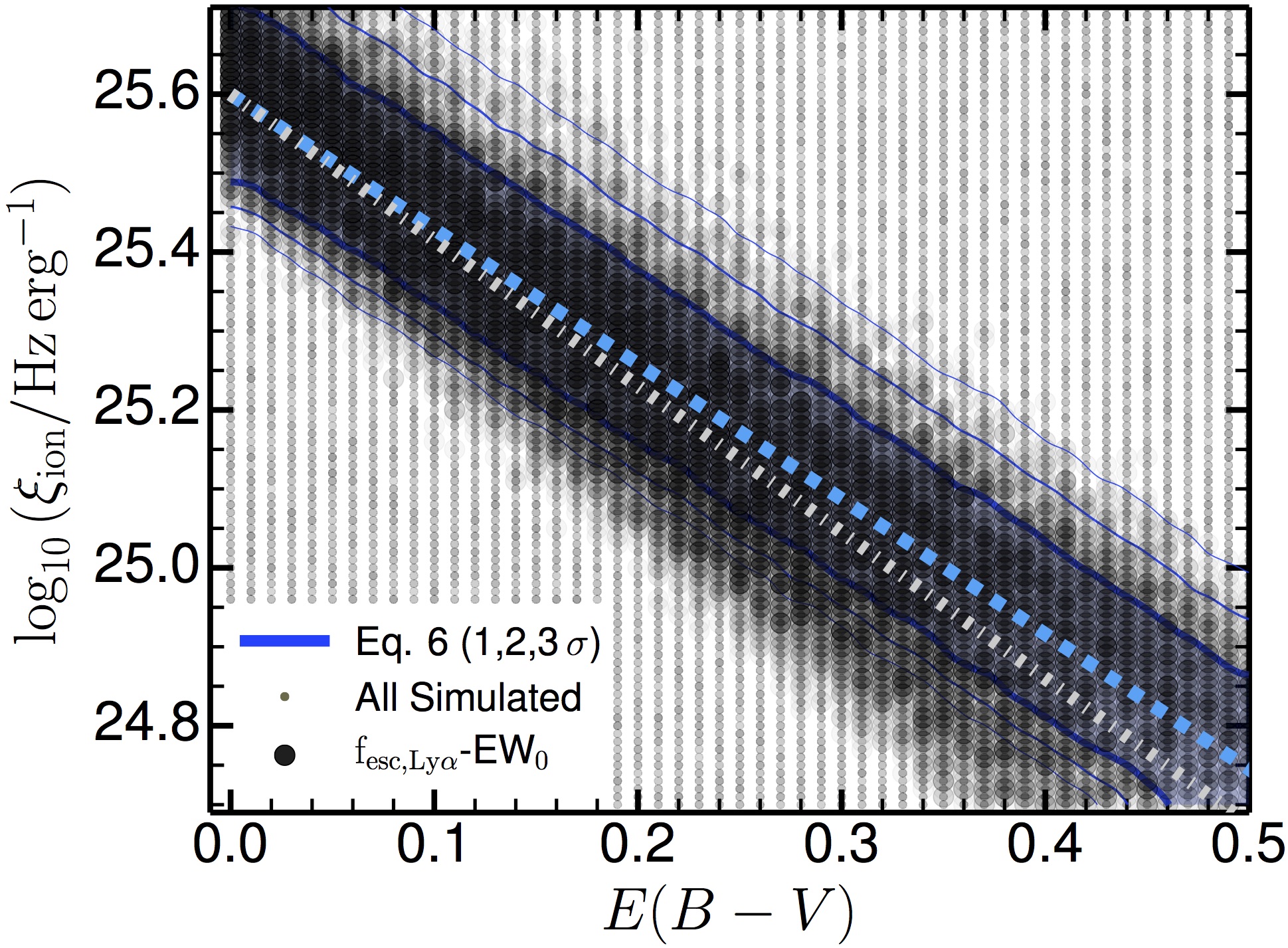}\\  
\end{tabular}
\caption{{\it Left:} The predicted relation between $\rm f_{esc,Ly\alpha}$ and Ly$\alpha$ EW$_0$ for our toy model, which shows little to no correlation by sampling all physical parameters independently (see Table \ref{Params_toy_model}). We also show the observed range ($\approx\pm0.05$) which is well constrained at $z\sim0-2.6$. We use simulated sources that are consistent with observations of LAEs to explore the potential reason behind the observed tight $\rm f_{esc,Ly\alpha}$-EW$_0$ correlation for LAEs. {\it Right:} By restricting our toy model to the observed relation and its scatter, we find a relatively tight $\rm \xi_{ion}$-$E(B-V)$ anti-correlation for LAEs (EW$_0>25$\,{\AA}): $\rm \log_{10}(\xi_{ion}/Hz\,erg^{-1})\approx-1.85\times\,E(B-V)+25.6$ (shown as grey dot-dashed line). This is in good agreement with the family of best fits using Equation \ref{This_study_DW10plus} (we show the 1, 2 and 3\,$\sigma$ contours) which yields $\rm \log_{10}(\xi_{ion}/Hz\,erg^{-1})\approx-1.71\times\,E(B-V)+25.6$, with only a small difference in the slope. The highest observed EW$_0$ correspond to the highest $\rm \xi_{ion}$ and the lowest $E(B-V)$, while lower EW$_0$ leads to a lower $\rm \xi_{ion}$ and a higher $E(B-V)$. Our results thus show that the observed $\rm f_{esc,Ly\alpha}$-EW$_0$ correlation for LAEs at $z\sim0-2.6$ only allows a well defined $\xi_{ion}$-$E(B-V)$ sequence that may be related with important physics such as the age of the stellar populations, their metallicity, dust production and how those evolve together.}
\label{Model_further_interp}
\end{figure*} 
%%%%%%%%%%%%%%%%%%%%%%%%%%%%

\subsection{The $\rm f_{esc,Ly\alpha}$-EW$_0$ and a potential $\xi_{ion}$-$E(B-V)$ sequence for LAEs}\label{Toy_model_interp}

We use our simple analytical model to further interpret the observed relation between $\rm f_{esc,Ly\alpha}$-EW$_0$ and its tightness. We take all artificially generated sources and select those that satisfy the observed relation given in Equation \ref{EW_fesc}, including its scatter (see Figure \ref{Model_further_interp}). We further restrict the sample to sources with Ly$\alpha$ EW$_0>25$\,{\AA}. We find that along the observed $\rm f_{esc,Ly\alpha}$-EW$_0$ relation, LAEs become less affected by dust extinction as a function of increasing EW$_0$, while $\rm \xi_{ion}$ increases, as already shown in \S\ref{phy_interp} and Figure \ref{EW_vs_f_esc_interpret}.

In the right panel of Figure \ref{Model_further_interp} we show the full parameter range explored in $\rm \xi_{ion}$-$E(B-V)$. By constraining the simulated sources with the observed $\rm f_{esc,Ly\alpha}$-EW$_0$ relation, we obtain a tight ($\pm0.1$\,dex), linear relation between $\rm \log_{10}\xi_{ion}$ and $E(B-V)$ given by $\rm \log_{10}(\xi_{ion}/Hz\,erg^{-1})\approx-1.85\times\,E(B-V)+25.6$. This means that in order for simulated sources to reproduce observations, LAEs should follow a very well defined $\rm \xi_{ion}$-$E(B-V)$ sequence with high $\rm \xi_{ion}$ values corresponding to very low $E(B-V)$ (mostly at high EW$_0$ and high $\rm f_{esc,Ly\alpha}$) and higher $E(B-V)$ to lower $\rm \xi_{ion}$ (mostly at low EW$_0$ and high $\rm f_{esc,Ly\alpha}$). Our results thus hint for the $\rm f_{esc,Ly\alpha}$-EW$_0$ to be driven by the physics (and diversity) of young and metal poor stellar populations and their evolution.

\subsection{Steps and equations for the model grid}\label{equations_grid}

We produce a model grid with our simple toy model which implements all equations and follows the observationally-motivated methodology used in the paper for full self-consistency. For each of the $N=1,000,000$ realisations, the script randomly picks (with a flat prior) parameters out of the parameter grid presented in Table \ref{Params_toy_model} (independently, per parameter).

The following steps are then taken per realisation. The H$\alpha$ luminosity is computed using the \cite{Kennicutt1998} calibration and the Ly$\alpha$ luminosity is obtained by using the case B coefficient used for that specific realisation. The UV SFR is computed by using $\rm \log_{10}(\xi_{ion}/Hz\,erg^{-1})$ for that realisation and the \cite{Kennicutt1998} calibration, which is then used to compute the intrinsic UV luminosity at rest-frame 1600\,{\AA} ($M_{UV}$ and L$_{UV}$). This step produces all the intrinsic luminosities which will be used: Ly$\alpha$, UV and H$\alpha$.

Next, by using the randomly picked value of $E(B-V)$ (see Table \ref{Params_toy_model}), the \cite{Calzetti00} attenuation law is used. For simplicity, as mentioned before, we set the attenuation of the nebular lines to be the same as the stellar continuum. We use the \cite{Calzetti00} dust attenuation law to compute $A_{\lambda}$ (mag) for $\lambda=1215.7,1600,6563$\,{\AA} in order to compute the attenuation at Ly$\alpha$, UV and H$\alpha$. We then compute the observed Ly$\alpha$, UV and H$\alpha$ luminosities after dust attenuation by computing:
\begin{equation} \label{obs_dust}
\rm L_{\lambda,observed}=L_{\lambda,intrinsic}\times10^{-0.4\,A_{\lambda}}
\end{equation}
Finally, for Ly$\alpha$, we apply the parameter ``Extra $\rm f_{esc,Ly\alpha}$" (see Table \ref{Params_toy_model}) which is multiplied by the observed Ly$\alpha$ luminosity (attenuated by dust) to produce the final observed Ly$\alpha$ luminosity. This is to quantify our ignorance on radiative transfer effects which are not explicitly modelled and are extremely complex. Following the methodology in this paper, the Ly$\alpha$ escape fraction is then computed using equation \ref{eq1} and with all quantities computed or randomly picked with the script.

Finally, after randomly picking an intrinsic $\beta_{int}$ slope, the \cite{Meurer1999} relation is used to transform $E(B-V)$ into an observed $\beta$ UV slope. This follows \cite{Meurer1999} and assumes that LAEs have $\beta=\beta_{int}$ for $E(B-V)=0.0$. $\beta$ is then used together with the observed UV luminosity at 1600\,{\AA} to compute the observed UV luminosity at $\lambda=1215.7$\,{\AA}. This is used to compute the observed EW$_0$. The toy model also computes the intrinsic EW$_0$, i.e., the rest-frame Ly$\alpha$ EW in the case of no dust and no scattering. The script also applies the calibrations derived/obtained or used in the paper to predict the Ly$\alpha$ escape fraction, Ly$\alpha$ and UV SFRs based on equations \ref{EW_fesc} and \ref{Lya_SFR_this} (see also Section \ref{highz_appli}) and the input from the simulation grid. These may be interesting for readers to explore further trends, and are provided as further information in the catalogue of 1,000,000 simulated sources.

\section{Data used for the high-redshift comparison between UV and Ly$\alpha$ SFRs}\label{Table_data}

%%%%%%%%%%%%%%%%%%%%%%%%%%%%%%%%%%%%%%%%%%%
%
% Table B1 - Application to high-z
%
%%%%%%%%%%%%%%%%%%%%%%%%%%%%%%%%%%%%%%%%%%%
\begin{table*}
 \centering
\caption{Application to high redshift UV-continuum and Ly$\alpha$ selected LAEs (see compilation by \citealt{Matthee2017followup}). Errors on Ly$\alpha$ luminosity and EW$_0$ are assumed to be $\approx0.1$\,dex, while errors on $\rm M_{UV}$ are taken as $\approx0.2$\,dex. We compute the UV SFRs (SFR$_{\rm UV}$, dust corrected) using \cite{Kennicutt1998} and $\beta$ values, individually measured when available, or $\beta=-1.6\pm0.2$ for UV-selected and $\beta=-1.9\pm0.2$ for Ly$\alpha$ selected sources when not available. Furthermore, we also compute dust corrected UV SFRs by using $\beta=-1.6\pm0.2$ for all sources (SFR$^{\beta=-1.6}_{\rm UV}$). Ly$\alpha$ SFRs (SFR$_{\rm Ly\alpha}$; calibrated to be dust-corrected) are computed with our Equation \ref{Lya_SFR_this}. Notes: 1: EW$_0$ have been recomputed and rest-framed when compared to original reference. 2: $\rm M_{UV}$ have been recomputed when compared to original reference. 3: Values used are from \cite{Zabl2015}. 4: Computed as in \cite{Matthee_CR7_17}. 5: COLA1's discovery is reported in \cite{Hu2016}; here we use the latest measurements from \cite{MattheeCOLA1}. 6: $\beta$ values not available; calculated assuming $\beta=-1.6\pm0.2$. 7: $\beta$ values not available; calculated assuming $\beta=-1.9\pm0.2$. This table is also provided in {\sc fits} format.}
\begin{tabular}{cccccccccc}
\hline
Name   &  $z$  & $\rm log_{10}(L_{Ly\alpha})$  & EW$_0$  & M$_{\rm UV}$  & $\beta_{\rm UV}$ & SFR$^{\beta}_{\rm UV}$ &  SFR$^{\beta=-1.6}_{\rm UV}$ & SFR$_{\rm Ly\alpha}$   &  Reference \\ 
(UV selected)  &    &  [erg\,s$^{-1}$] & [{\AA}]  &  [mag]  &   & [M$_{\odot}$\,yr$^{-1}$] &  [M$_{\odot}$\,yr$^{-1}$] &  [M$_{\odot}$\,yr$^{-1}$]  &   \\
\hline
  A383-5.2  & 6.03  & $42.8$ &  $138$   &  $-19.3$   & $\beta^6$ &  $10^{+5}_{-3}$ &  $10^{+3}_{-3}$  &  $11^{+4}_{-3}$  & \cite{Stark2015} \\
 RXCJ22-ID3  & 6.11  & $42.5$ &  $40$   &  $-20.1$   & $-2.3\pm0.2$ &  $7^{+2}_{-1}$ &  $21^{+7}_{-5}$  &  $16^{+8}_{-5}$  & \cite{Mainali2017} \\
 RXCJ22-4431  & 6.11  & $42.9$ &  $68$   &  $-20.2$   & $-2.3\pm0.2$ &  $8^{+2}_{-1}$ &  $23^{+8}_{-6}$  &  $25^{+11}_{-7}$  & \cite{Schmidt2017} \\
 SDF-46975  & 6.84  & $43.2$ &  $43$   &  $-21.5$   & $\beta^6$ &  $76^{+37}_{-25}$ &  $76^{+26}_{-20}$  &  $75^{+35}_{-22}$  & \cite{Ono2012} \\
 IOK-1  & 6.96  & $43.0$ &  $42$   &  $-21.3$   & $-2.0\pm0.3$ &  $31^{+15}_{-9}$ &  $63^{+22}_{-16}$  &  $57^{+26}_{-17}$  & \cite{Ono2012} \\
 BDF-521  & 7.01  & $43.0$ &  $64$   &  $-20.6$   & $-2.3\pm0.4$ &  $11^{+3}_{-2}$ &  $33^{+11}_{-9}$  &  $34^{+14}_{-10}$  & \cite{Cai2015} \\
 A1703 zd6  & 7.04  & $42.5$ &  $65$   &  $-19.3$   & $-2.4\pm0.2$ &  $3^{+1}_{-1}$ &  $10^{+3}_{-3}$  &  $10^{+4}_{-3}$  & \cite{Stark2015b} \\
 BDF-3299  & 7.11  & $42.8$ &  $50$   &  $-20.6$   & $-2.0\pm0.5$ &  $16^{+8}_{-5}$ &  $33^{+11}_{-9}$  &  $30^{+13}_{-9}$  & \cite{Vanzella2011} \\
 GLASS-stack  & 7.20  & $43.0$ &  $210$   &  $-19.7$   & $\beta^6$ &  $15^{+7}_{-5}$ &  $15^{+5}_{-4}$  &  $10^{+4}_{-3}$  & \cite{Smidt2016} \\
 EGS-zs8-2  & 7.48  & $42.7$ &  $9$   &  $-21.9$   & $-1.8\pm0.4$ &  $77^{+37}_{-25}$ &  $110^{+37}_{-28}$  &  $102^{+205}_{-49}$  & \cite{Stark2015a} \\
 FIGSGN1-1292  & 7.51  & $42.8$ &  $49$   &  $-21.2$   & $\beta^6$ &  $58^{+29}_{-19}$ &  $58^{+20}_{-15}$  &  $31^{+14}_{-9}$  & \cite{Tilvi2016} \\
 GN-108036  & 7.21  & $43.2$ &  $33$   &  $-21.8$   & $\beta^6$ &  $100^{+52}_{-33}$ &  $100^{+35}_{-26}$  &  $99^{+49}_{-31}$  & \cite{Stark2015a} \\
 EGS-zs8-1  & 7.73  & $43.1$ &  $21$   &  $-22.1$   & $-1.7\pm0.1$ &  $111^{+56}_{-37}$ &  $132^{+46}_{-34}$  &  $125^{+83}_{-43}$  & \cite{Oesch2015} \\
\hline
(Ly$\alpha$ selected)  &    &   &  &    &   &  &   \\
\hline
 SR6$^1$  & 5.68  & $43.4$ &  $210$   &  $-21.1$   & $-1.8\pm0.4$ &  $38^{+19}_{-13}$ &  $53^{+17}_{-14}$  &  $26^{+10}_{-7}$  & \cite{Matthee2017followup} \\
 Ding-3  & 5.69  & $42.8$ &  $62$   &  $-20.9$   & $\beta^7$ &  $25^{+13}_{-8}$ &  $44^{+15}_{-11}$  &  $24^{+10}_{-7}$  & \cite{Ding2017} \\
 Ding-4  & 5.69  & $42.3$ &  $106$   &  $-20.5$   & $\beta^7$ &  $18^{+9}_{-6}$ &  $30^{+11}_{-7}$  &  $4^{+2}_{-1}$  & \cite{Ding2017} \\
 Ding-5  & 5.69  & $43.2$ &  $79$   &  $-21.0$   & $\beta^7$ &  $28^{+14}_{-9}$ &  $48^{+17}_{-12}$  &  $44^{+18}_{-12}$  & \cite{Ouchi2008} \\
 Ding-1  & 5.70  & $43.0$ &  $21$   &  $-22.2$   & $\beta^7$ &  $85^{+43}_{-28}$ &  $147^{+50}_{-38}$  &  $103^{+72}_{-35}$  & \cite{Ding2017} \\
 J233454$^2$  & 5.73  & $43.7$ &  $210$   &  $-21.5$   & $\beta^7$ &  $44^{+22}_{-14}$ &  $76^{+27}_{-19}$  &  $50^{+20}_{-14}$  & \cite{Shibuya2017_spec} \\
 J021835  & 5.76  & $43.7$ &  $107$   &  $-21.7$   & $\beta^7$ &  $53^{+26}_{-17}$ &  $92^{+31}_{-24}$  &  $94^{+39}_{-27}$  & \cite{Shibuya2017_spec} \\
 VR7$^1$  & 6.53  & $43.4$ &  $35$   &  $-22.5$   & $-2.0\pm0.3$ &  $97^{+48}_{-31}$ &  $192^{+65}_{-48}$  &  $149^{+72}_{-46}$  & \cite{Matthee2017followup} \\
 J162126$^2$  & 6.55  & $43.9$ &  $99$   &  $-22.8$   & $\beta^7$ &  $145^{+73}_{-48}$ &  $252^{+88}_{-64}$  &  $170^{+68}_{-48}$  & \cite{Shibuya2017_spec} \\
 J160234  & 6.58  & $43.5$ &  $81$   &  $-21.9$   & $\beta^7$ &  $63^{+31}_{-21}$ &  $111^{+37}_{-28}$  &  $88^{+36}_{-25}$  & \cite{Shibuya2017_spec} \\
 Himiko$^3$  & 6.59  & $43.6$ &  $65$   &  $-22.1$   & $-2.0\pm0.4$ &  $63^{+32}_{-19}$ &  $132^{+45}_{-33}$  &  $142^{+58}_{-42}$  & \cite{Ouchi2009} \\
 COLA1  & 6.59  & $43.6$ &  $120$   &  $-21.6$   & $\beta^7$ &  $48^{+24}_{-16}$ &  $83^{+29}_{-22}$  &  $73^{+29}_{-21}$  & \cite{MattheeCOLA1} \\
 CR7$^4$  & 6.60  & $43.9$ &  $211$   &  $-22.2$   & $-2.3\pm0.3$ &  $49^{+15}_{-9}$ &  $146^{+50}_{-37}$  &  $87^{+35}_{-25}$  & \cite{Sobral2015} \\

 \hline
\end{tabular} \label{HIghz_appl_tabl}
\end{table*}

Table \ref{HIghz_appl_tabl} provides the data used for Figure \ref{SFR_vs_SFR}, including individual measurements per source, their name and reference. Note that the data is taken from a compilation from \cite{Matthee2017followup} with minor modifications for a few LAEs, as a indicated in Table \ref{HIghz_appl_tabl}.

\end{appendix}

\end{document}